\documentclass[10pt, conference]{IEEEtran}

\newif\iflong
\longtrue

\usepackage{subcaption}
\usepackage{xcolor}
\definecolor{darkblue}{rgb}{0,0,0.4}

\definecolor{colorstrong}{HTML}{33A02C}
\definecolor{colormedium}{HTML}{A6611A}
\definecolor{colorweak}{HTML}{1F78B4}
\definecolor{colorarrow}{HTML}{000000}

\usepackage{hyperref}
\hypersetup{
  colorlinks,
  bookmarks,
  plainpages=false,
  pdfpagelabels,
  linkcolor=darkblue,
  citecolor=darkblue,
  pagecolor=darkblue,
  urlcolor=darkblue,
  pdfstartview=FitH,
  pdfauthor={C.A. Furia},
  pdftitle={Bayesian Statistics in Software Engineering}
}
\usepackage{url}

\makeatletter
\renewcommand*{\thetable}{\arabic{table}}
\renewcommand*{\thefigure}{\arabic{figure}}
\let\c@table\c@figure
\makeatother

\usepackage{xspace}
\newcommand{\rosetta}{Rosetta\xspace}
\newcommand{\bench}{Bench\xspace}
\newcommand{\agile}{AvsS\xspace}
\newcommand{\itproj}{ITP\xspace}
\newcommand{\testing}{ST\xspace}

\newcommand{\rqu}[1]{RQ$_{\text{#1}}$}
\newcommand{\ans}[1]{AN$_{\text{#1}}$}

\usepackage[T1]{fontenc}
\usepackage[scaled=0.81]{beramono} 
\usepackage[fleqn,cmex10]{amsmath}
\renewcommand{\vec}[1]{\mathbf{#1}}
\newcommand{\bvec}[1]{\mathbf{\b{#1}}}
\renewcommand{\b}[1]{\overline{#1}}
\DeclareMathOperator{\sgn}{sgn}
\DeclareMathOperator{\argmin}{argmin}
\newcommand{\outcome}{\omicron}
\newcommand{\stakeholders}{\iota}
\usepackage{mathptmx}
\newcommand{\pmf}{p.m.f.\xspace}
\newcommand{\pdf}{p.d.f.\xspace}
\newcommand{\cdf}{c.d.f.\xspace}
\newcommand{\pall}{p_{*}}
\newcommand{\pbetter}{p_{\mathbb{A}}}
\newcommand{\pworse}{p_{\mathbb{S}}}
\newcommand{\original}{t}
\newcommand{\voriginal}{\vec{\original}}
\newcommand{\mbc}{T}
\newcommand{\vmbc}{\vec{\mbc}}
\newcommand{\na}{\widehat{\alpha}}
\newcommand{\nb}{\widehat{\beta}}
\newcommand{\correct}{\textsc{ok}}
\newcommand{\wrong}{\textsc{err}}

\usepackage{amsfonts}
\newcommand{\prob}{\mathbb{P}}
\newcommand{\like}{\mathcal{L}}
\newcommand{\prior}{\pi}
\newcommand{\post}{\mathcal{P}}

\usepackage[font={scriptsize,normal},skip=5pt]{caption}
\usepackage{array}

\usepackage{colortbl}
\usepackage{enumerate}

\let\labelindent\relax 
\usepackage[inline]{enumitem}

\newcommand{\fakepar}[1]{\textbf{#1.}}
\newcommand{\fakeparnp}[1]{\textbf{#1}}

\newcommand{\notsig}[1]{\cellcolor[RGB]{96,96,96}{#1}}
\newcommand{\weaksig}[1]{\cellcolor[RGB]{160,160,160}{#1}}
\newcommand{\strongsig}[1]{#1}

\usepackage{changepage}

\usepackage[colorinlistoftodos,textsize=tiny]{todonotes}

\usepackage{graphicx}
\usepackage{grffile} 
\newcommand{\includegraphicsifexists}[2][,]{\IfFileExists{#2}{\includegraphics[#1]{#2}}{}}

\usepackage{tikz}
\usetikzlibrary{shapes,arrows,positioning,calc}

\usepackage{boxedminipage}
{\smallskip
\noindent
\let\emph=\textbf
\begin{boxedminipage}{\columnwidth}\begin{center}\em}%
{\end{center}\end{boxedminipage}%
\smallskip
}

\newenvironment{boxcenter}[1][\topsep]
  {\setlength{\topsep}{#1}\par\kern\topsep\centering}
  {\par\kern\topsep}
\usepackage{fancybox}
\newcommand{\takehome}[1]{\begin{boxcenter}[1pt]
\ovalbox{
    \begin{minipage}{0.95\columnwidth}
      {\it #1}
    \end{minipage}
}\end{boxcenter}
}
\newcommand{\rquestion}[1]{\begin{boxcenter}[1pt]
\fbox{
    \begin{minipage}{0.95\columnwidth}
      {#1}
    \end{minipage}
}\end{boxcenter}
}

\begin{document}

\title{What Would Bayes Do? Bayesian Data Analysis in Software Engineering}
\title{Bayesian Data Analysis in Software Engineering: A Practical Guide}
\title{Bayesian Data Analysis in Software Engineering}
\title{Bayesian Statistics in Software Engineering}
\title{Bayesian Statistics in Software Engineering: Practical Guide and Case Studies}

\author{\IEEEauthorblockN{Carlo A.\ Furia}
\IEEEauthorblockA{Chalmers University of Technology, Gothenburg, Sweden\\
\url{furia@chalmers.se}$\quad$\url{bugcounting.net}}
}

\maketitle

\begin{abstract}
Statistics comes in two main flavors: frequentist and Bayesian.
For historical and technical reasons, frequentist statistics has dominated data analysis in the past; but Bayesian statistics is making a comeback at the forefront of science. 
In this paper, we give a practical overview of Bayesian statistics and illustrate its main advantages over frequentist statistics for the kinds of analyses that are common in empirical software engineering, where frequentist statistics still is standard.
We also apply Bayesian statistics to empirical data from previous research investigating agile vs.\ structured development processes, the performance of programming languages, and random testing of object-oriented programs.
In addition to being case studies demonstrating how Bayesian analysis can be applied in practice, they provide insights beyond the results in the original publications (which used frequentist statistics), thus showing the practical value brought by Bayesian statistics.
\end{abstract}

\section{Introduction}
\label{sec:introduction}

A towering figure in evolutionary biology and statistics, Ronald Fisher has exerted a tremendous influence on pretty much all of experimental science since the early 20th century.
The statistical techniques he developed or perfected constitute the customary data analysis toolset of \emph{frequentist statistics}, 
in direct contrast to the other school of statistics---known as \emph{Bayesian} since it is ultimately based on Bayes theorem of conditional probabilities.
The overwhelming prevalence of frequentist statistics in all the sciences was due partly to Fisher's standing and keen efforts of promotion, partly to its claim of being ``more objective'', and partly to its techniques being less computationally demanding than Bayesian ones---a crucial concern with the limited computational resources available in the first part of the past century.

In the last couple of decades, however, the scientific community has begun a critical re-examination of the toolset of frequentist statistics, with particular focus on the widespread technique of statistical hypothesis testing using $p$-values.
The critics\iflong, who include prominent statisticians such as Cohen~\cite{cohen-effectsizes},\fi{} have observed methodological shortcomings~\cite{pvalue-cohen,all-false} 
and, more generally, limitations of the frequentist's rigid view.
On the other hand, the difficulties of applying Bayesian analysis due to its higher computational demands have become moot with the vast computing power available nowadays.
As a result, Bayesian techniques are becoming increasingly popular, and have buttressed spectacular advances in automation and machine learning such as the deep neural networks that powered Google's AlphaGo~\cite{alphago}.

As we argue in \autoref{sec:relatedwork}, frequentist statistics is still dominant in empirical software engineering research, whose best practices have been perfected later than in other experimental sciences.
The main contribution of this paper is thus \emph{casting the usage of Bayesian statistics as an alternative and as a supplement to frequentist statistics} in the context of the \emph{data analyses} that are common in \emph{software engineering}.

To make the presentation self contained, in \autoref{sec:bayes-stats} we briefly recall some fundamental notions of probability, and then introduce Bayes theorem---the cornerstone of Bayesian analysis.
In \autoref{sec:classic-vs-bayes} we explain the shortcomings of frequentist statistical hypothesis testing, and suggest Bayes factors as an alternative technique.
We also present other analyses that are fueled by Bayes theorem, and argue about the significant advantages of taking a Bayesian point of view.

In \autoref{sec:case-studies} we then proceed to ``eat our own dog food'' and demonstrate Bayesian analysis on \emph{three case studies} whose main data is taken from previous work of ours on 
agile vs.\ structured development processes~\cite{agile,agile-extended}, the performance of programming languages~\cite{rosetta}, and random testing with specifications~\cite{testing}.
We focused on some of our own publications both because their data was (obviously) more readily available to us, and to show that we too used to rely entirely on the toolset of frequentist analysis.
In each case, we take the same data that we analyzed using frequentist statistics in the original publications and, after briefly summarizing the original analysis and its results, we describe a new analysis that refines the original results or increases the confidence we can have in them.
In two case studies we also supplement the original experiments with additional data obtained by other researchers in comparable conditions.
This turns on its head the criticism that Bayesian analysis is ``less objective'' than frequentist one because it depends on prior information: incorporating independently obtained information can be, in fact, conducive to richer and more robust analyses---provided it is done sensibly following justifiable modeling choices.

\autoref{sec:threats} discusses threats to validity, emphasizing where Bayesian statistics can help mitigate them.
\autoref{sec:conclusions} concludes with practical guidelines to applying Bayesian analysis in empirical software engineering.
Researchers should be familiar with all the possibilities offered by statistics and able to deploy the best tools of the trade pragmatically in each situation.
Since frequentist techniques are already well understood, it is time to make some room for Bayesian analysis.

\fakepar{Extended version}
\autoref{sec:case-studies}'s case studies focus on the design of the analyses and their main results; \iflong{} the appendix \else{} a technical report~\cite{bayes-extended}\fi{} includes more details about measures and plots.

\section{Related Work}
\label{sec:relatedwork}

\fakepar{Empirical research in software engineering}
Statistical analysis of empirical data has become commonplace in software engineering research~\cite{experiments-book}, and it is even making its way into software development practices~\cite{data-scientist}.
As we discuss below, the overwhelming majority of statistical techniques that are being used in software engineering empirical research are, however, of the frequentist kind, with Bayesian statistics hardly even mentioned.
Of course, Bayesian statistics are a fundamental component of many machine learning techniques~\cite{ml-springer,ml-book}; as such, they are used in software engineering research indirectly whenever machine learning is used.
In this paper, however, we are concerned with the direct usage of statistics on empirical data, which is where the state of the art in software engineering seems mainly confined to frequentist techniques.
\iflong
As we argue in the rest of the paper, this is a lost opportunity because Bayesian techniques do not suffer from some technical limitations of frequentist ones, and can support rich, robust analyses in several situations.
\fi

\fakeparnp{Bayesian analysis in software engineering?}
To validate the perception that Bayesian statistics are not normally used in empirical software engineering, we carried out a small literature review of ICSE papers.\iflong\footnote{\cite{views-icse15} has a much more extensive literature survey of empirical publications in software engineering.}\fi{} 
We selected all papers from the latest four editions of the International Conference on Software Engineering (ICSE 2013 to ICSE 2016) that mention ``empirical'' in their title or in their section's name in the proceedings.
This gave 22 papers, 
from which we discarded one~\cite{stochastic-icse14} that is actually not an empirical study.
The experimental data in the remaining 21 papers come from various sources: 
the output of analyzers and other programs~\cite{browser-icse13,configuration-icse14,equivalence-icse15,compilers-icse16}, 
the mining of repositories of software and other artifacts~\cite{fixing-icse13,evolving-icse14,fixes-icse15,verification-icse15,js-icse16}, 
the outcome of controlled experiments involving human subjects~\cite{smells-icse13,lambdas-icse16,summaries-icse16},
interviews and surveys~\cite{uml-icse13,codereviews-icse13,coupling-icse13,views-icse15,belief-icse16,green-icse16,data-icse16},
and a literature review~\cite{grounded-icse16}.

As one would expect from a top-tier venue like ICSE, the papers follow the recommended practices in reporting and analyzing data at least to some extent, using significance testing (5 papers), effect sizes (3 papers), correlation coefficients (4 papers), frequentist regression (2 papers), and visualization in charts or tables (20 papers).
None of the papers, however, uses Bayesian statistics.
In fact, no paper but two~\cite{fixing-icse13,belief-icse16} even mentions the terms ``Bayes'' or ``Bayesian''.
One exception~\cite{fixing-icse13} only cites  Bayesian machine-learning techniques used in related work to which it compares.
The other exception~\cite{belief-icse16} includes a presentation of the two views of frequentist and Bayesian statistics---with a critique of $p$-values similar to the one we make in \autoref{sec:classic-vs-bayes}---but does not show how the latter can be used in practice.
\cite{belief-icse16}'s main aim is investigating the relationship between empirical findings in software engineering and the actual beliefs of programmers about the same topics.
To this end, \iflong{} it is based on a survey of programmers whose responses are analyzed using frequentist statistics;\fi{} Bayesian statistics is mentioned to frame the discussion about the relationship between evidence and beliefs\iflong{} (but it is not mentioned after the introductory second section)\fi{}.
Our paper has a more direct aim: concretely showing how Bayesian analysis can be applied in practice in empirical software engineering research\iflong, as an alternative to frequentist statistics\fi; thus, its scope is largely complementary to \cite{belief-icse16}'s.

\fakepar{Criticism of the $p$-value}
Statistical hypothesis testing---and its summary outcome, the $p$-value---has been customary in experimental science for many decades, both for the influence of his proponents Fisher, Neyman, and Pearson, and because it offers a straightforward, ready-made procedure that is computationally simple.
More recently, criticism of frequentist hypothesis testing has been voiced in many experimental sciences, such as psychology~\cite{pvalue-psychology,pvalue-cohen} and medicine~\cite{pvalue-medicine}, that used to rely on it heavily, as well as in statistics research itself~\cite{pvalue-statisticians,gelman-pvalues}.
The criticism, which we articulate in \autoref{sec:classic-vs-bayes}, concludes that $p$-value-based hypothesis testing should be abandoned.
There has been no similar explicit criticism of $p$-values in software engineering research, and in fact statistical hypothesis testing is still regularly used.

\fakepar{Guidelines for using statistics}
Best practices of using statistics in empirical software engineering are described in a few books~\cite{experiments-book,datascience-perspectives} and articles~\cite{hitchhiker-icse11,hitchhiker-journal,JedlitschkaJR14}.
Given their focus on frequentist statistics,\footnote{\cite{hitchhiker-icse11,experiments-book,JedlitschkaJR14} do not mention Bayesian techniques; \cite{hitchhiker-journal} mentions them only to declare they are not discussed; one chapter~\cite{bayes-in-book} of~\cite{datascience-perspectives} outlines Bayesian networks as a machine learning technique.} they all are complementary to the present paper, whose main goal is showing how Bayesian techniques can add to, or replace, frequentist ones, and how they can be applied in practice.

\section{A Practical Overview of Bayesian Statistics}
\label{sec:bayes-stats}

Statistics provides models of \emph{events}, such as the output of a randomized algorithm; the probability function $\prob$ assigns probabilities---values in the real unit interval $[0, 1]$, or equivalently percentages in $[0, 100]$---to events.
Often, events are values taken by \emph{random variables} that follow a certain probability \emph{distribution}.
For example, if $X$ is a random variable modeling the throwing of a six-face dice, it means that $\prob[x] = 1/6$ for $x \in [1..6]$, and $\prob[x] = 0$ for $x \not\in [1..6]$---where $\prob[x]$ is a shorthand for $\prob[X = x]$, and $[m..n]$ is the set of integers between $m$ and $n$.

The probability of variables over discrete domains is described by \emph{probability mass functions} (\pmf for short); their counterparts over continuous domains are probability density functions (\pdf), whose integrals give probabilities.
In this paper we mostly deal with discrete domains and \pmf, or \pmf approximating \pdf, although most notions apply to continuous-domain variables as well with a few technical differences.
For convenience, we may denote a distribution and its \pmf with the same symbol; for example, random variable $X$ has a \pmf also denoted $X$, such that $X[x] = \prob[x] = \prob[X = x]$.

\fakepar{Conditional probability}
The \emph{conditional} probability \linebreak $\prob[h \mid d]$ is the probability of $h$ given that $d$ has occurred.
When modeling experiments, $d$ is the empirical data that has been recorded, and $h$ is a hypothesis that is being tested.
Consider a static analyzer that outputs $\top$ (resp.\ $\bot$) to indicate that the input program never overflows (resp.\ may overflow); $\prob[\correct \mid \top]$ is the probability that, when the algorithm outputs $\top$, the input is indeed free from overflows---the data is the output ``$\top$'' and the hypothesis is ``the input does not overflow''.

\subsection{Bayes Theorem}
\label{sec:bayes-theorem}

\fakeparnp{Bayes theorem} 
connects the conditional probabilities \linebreak$\prob[h \mid d]$ and $\prob[d \mid h]$:
\begin{equation}
\prob[h \mid d]
\ =\ 
\frac{\prob[d \mid h]\cdot\prob[h]}{\prob[d]}\,.
\label{eq:bayes-th}
\end{equation}
Suppose that the static analyzer gives true positives and true negatives with high probability ($\prob[\top \mid \correct] = \prob[\bot \mid \wrong] = 0.99$), and that many programs are affected by some overflow errors ($\prob[\correct] = 0.01$).
Whenever the analyzer outputs $\top$, what is the chance that the input is indeed free from overflows?
Using Bayes theorem, $\prob[\correct \mid \top] = (\prob[\top \mid \correct]\, \prob[\correct])/\prob[\top] = (\prob[\top \mid \correct]\, \prob[\correct])/(\prob[\top \mid \correct]\, \prob[\correct] + \prob[\top \mid \wrong]\, \prob[\wrong]) = (0.99 \cdot 0.01)/(0.99 \cdot 0.01 + 0.01 \cdot 0.99) = 0.5$, we conclude that we can have a mere 50\% confidence in the analyzer's output.

\fakepar{Priors, likelihoods, and posteriors}
In Bayesian analysis~\cite{thinkBayes}, each factor of \eqref{eq:bayes-th} has a special name:
\begin{enumerate*}
\item $\prob[h]$ is the \emph{prior}---the probability of the hypothesis before having considered the data---written $\prior[h]$;
\item $\prob[d \mid h]$ is the \emph{likelihood} of the data under the hypothesis---written $\like[d; h]$;
\item $\prob[d]$ is the \emph{normalizing constant};
\item and $\prob[h \mid d]$ is the \emph{posterior}---the probability of the hypothesis after taking the data into account---written $\post_d[h]$.
\end{enumerate*}
With this terminology, we say that \emph{the posterior is proportional to the likelihood times the prior}.

The only role of the normalizing constant is ensuring that the posterior defines a correct probability distribution when evaluated over all hypotheses.
In most cases we deal with hypotheses $h \in H$ that are mutually exclusive and exhaustive; then, the normalizing constant is simply $\prob[d] = \sum_{h \in H} \prob[d \mid h]\,\prob[h]$, which can be computed from the rest of the information: we say that we \emph{update} the prior to get the posterior.
Thus, it normally suffices to define likelihoods that are \emph{proportional to} a probability, and rely on this \emph{update rule} to normalize them and get a proper probability distribution as posterior.

In case of repeated experiments, the data is a set $D$ that collects the outcomes of all experiments.
Bayes' update can be \emph{iterated}: update the prior to get the posterior $\post_{d_1}[h]$ using some $d_1 \in D$; then the posterior becomes the new prior, which is updated using $d_2 \in D$ to get a new posterior $\post_{d_2}[h]$; and so on for all $d \in D$.

\subsection{Frequentist vs.\ Bayesian Statistics}
\label{sec:classic-vs-bayes}

Despite being a simple result about an elementary fact in probability, Bayes theorem has significant implications in the way we can reason about statistics.
We do not discuss the philosophical differences between how frequentist and Bayesian statistics interpret their results.
Instead, we focus on describing how some features of Bayesian statistics support new ways of analyzing data.
\iflong
We start by criticizing statistical hypothesis testing since it is a customary technique in frequentist statistics that is widely applied in experimental science, and suggest how Bayesian techniques could provide more reliable analyses.
\fi
\autoref{sec:case-studies} will then demonstrate them in practice on significant case studies.

\fakepar{Hypothesis testing vs.\ model comparison}
A primary goal of experimental science is validating models of behavior based on empirical data.
This often takes the form of choosing between alternative hypotheses, such as deciding whether a programming language is faster than another (\autoref{sec:rosetta})\iflong, or whether agile development methods lead to more successful projects (\autoref{sec:agile})\fi{}.
\emph{Hypothesis testing} is the customary framework offered by frequentist statistics to choose between hypotheses.
In the classical setting, a null hypothesis $h_0$ corresponds to ``no significant difference'' between two \emph{treatments} $A$ and $B$ (such as two static analysis algorithms whose effectiveness we want to compare); an alternative hypothesis $h_1$ is the null hypothesis's negation, which corresponds to a significant difference between applying $A$ and applying $B$.
A \emph{statistical significance test}~\cite{hitchhiker-journal}, such as the $t$-test or the $U$-test, is a procedure that inputs two datasets $D_A$ and $D_B$, respectively recording the outcome of applying $A$ and $B$, and outputs a probability called the $p$-value.
The $p$-value is the likelihood of the data under the null hypothesis; namely, it is the conditional probability $\prob[D \mid h_0]$ that the outcomes in $D = D_A \cup D_B$ would occur assuming that the treatments $A$ and $B$ are equivalent (or, in more precise statistical terms, determine outcomes with the same distribution).
If the $p$-value is sufficiently small---typically $p \leq 0.05$ or $p \leq 0.01$---we \emph{reject} the null hypothesis, which corresponds to leaning towards preferring the alternative hypothesis $h_1$ over $h_0$: we have confidence that $A$ and $B$ differ.

Unfortunately, this widely used approach to testing hypotheses suffers from serious shortcomings.
The most glaring problem is that, in order to decide whether $h_0$ is a plausible explanation of the data, we would need the conditional probability $\prob[h_0 \mid D]$ of the hypothesis given the data, not the $p$-value $\prob[D \mid h_0]$.
The two conditionals probabilities are related by Bayes theorem \eqref{eq:bayes-th}, so knowing only $\prob[D \mid h_0]$ is not enough to determine $\prob[h_0 \mid D]$;\footnote{Assuming that they are equal is the ``confusion of the inverse''~\cite{fallacy-inverse}.} in fact, \autoref{sec:bayes-theorem}'s example of the static analyzer showed a case where one conditional probability is 99\% while the other is only 50\%.
Other problems come from how hypothesis testing pits the null hypothesis against the alternative hypothesis: as the number of observations grows, it becomes increasingly likely that \emph{some} effect is detectable (or, conversely, it becomes increasingly unlikely that \emph{no effects} are), which leads to rejecting the null hypothesis, independent of the alternative hypothesis, just because it is unreasonably restrictive.
This problem may result both in suggesting that some negligible effect is significant just because we reject the null hypothesis, and, conversely, in discarding some interesting experimental results just because they fail to trigger the $p \leq 0.05$ threshold of significance.
This is part of the more general problem with insisting on binary decisions between two alternatives: a better approach would be based on richer statistics than one or few summary values (such as the $p$-value) and would combine quantitative and qualitative data to get richer pictures.

In Bayesian statistics, the closest alternative to statistical significance testing is model comparison based on \emph{Bayes factors}.\footnote{Popularized by Jeffreys~\cite{jeffreys-book}, who developed it independent of Turing~\cite{good-origin-bfactor}.}
To evaluate whether a hypothesis $H_1$ is a better explanation of the data $D$ than another hypothesis $H_2$, we compute the factor $K(D) = \prob[D \mid H_1] / \prob[D \mid H_2]$, which corresponds to a ratio of likelihoods.
In Bayesian analysis, $H_1$ and $H_2$ normally are not two fixed hypotheses like $h_0$ and $h_1$, but two \emph{families} of hypotheses with associated probability distributions, so that we can compute the Bayes factor as the ratio of weighted sums:
\[
K(D) = 
\frac{\sum_{x \in H_1} \prob[x]\cdot\prob[D \mid x]}{\sum_{y \in H_2} \prob[y]\cdot\prob[D \mid y]}\,.
\]
The ratio of posteriors equals the Bayes factor times the ratio of priors, $\post_D(H_1)/\post_D(H_2) = K(D) \cdot \prior(H_1)/\prior(H_2)$; thus, $K(D)$ indicates \emph{how much the data is likely to shift the prior} belief towards $H_1$ over $H_2$.
Choosing between hypotheses based on Bayes factors avoids the main pitfalls of $p$-values---which capture information that is not conclusive.
Jeffreys~\cite{jeffreys-book} suggests the following scale to interpreting $K(D)$: 
\begin{center}
\scriptsize
\setlength{\tabcolsep}{2pt}
\begin{tabular}{rcl l}
 \multicolumn{3}{c}{} & \textsc{evidence for $H_1$} \\
\hline
      & $K(D)$ & $< \ \ 1$ & negative (supports $H_2$) \\
$1 <$ & $K(D)$ & $\leq\ \ 3$ & barely worth mentioning \\
$3 <$ & $K(D)$ & $\leq \ 10$ & substantial \\
$10 <$ & $K(D)$ & $\leq \ 32$ & strong \\
$32 <$ & $K(D)$ & $\leq 100$ & very strong \\
$100 <$ & $K(D)$ &  & decisive \\
\end{tabular}
\end{center}

\fakepar{Scalar summaries vs.\ posterior distributions}
Decisions based on Bayes factors still reduce statistical modeling to binary choices; but a distinctive advantage of full-fledged Bayesian statistics is that it supports deriving a complete \emph{distribution} of posterior probabilities, by applying \eqref{eq:bayes-th} for all hypotheses $h$, rather than just scalar summaries (such as estimators of mean, median, and standard deviation, or standardized measures of effect size).
Given a distribution we can still compute scalar summaries, but we retain additional advantages of being able to visualize the distribution, as well as to derive other distributions by iterative application of Bayes theorem.
This supports decisions based on a variety of criteria and on a richer understanding of the experimental data, as we demonstrate in the case studies of \autoref{sec:rosetta} and \autoref{sec:testing}.
\iflong
In \autoref{sec:rosetta}, for example, we visually inspect the posterior distribution to get an idea of whether some borderline differences in performance between programming languages can be considered significant; in \autoref{sec:testing}, we derive the distribution of all bugs in a module from the posterior of the bugs found by random testing in one module.
\fi

\fakepar{The role of prior information}
The other distinguishing feature of Bayesian analysis is that it \emph{starts from a prior probability} which models the initial knowledge about the hypotheses.
The prior can record previous results in a way that is congenial to the way science is supposed to work---not as completely independent experiments in a metaphorical vacuum, but by constantly scrutinizing previous results and updating our models based on new evidence.
A kind of canned criticism observes that using a prior is a potential source of bias.
However, explicitly taking into account this very fact helps analyses being more rigorous.
In particular, we can often consider several different alternative priors to perform Bayesian analysis.
Priors that do not reflect any strong assumptions are called \emph{uninformative}; a uniform distribution over hypotheses is the most common example.
If it turns out that the posterior distribution is largely independent of the chosen prior, we say that the data \emph{swamps} the prior, and hence the experimental evidence is quite strong.
If, conversely, choosing a suitable prior is necessary to get sensible results, it means that the evidence is not overwhelming, and hence any additional reliable source of information should be vetted and used to sharpen the analysis results.
Bayesian analysis stresses the importance of careful \emph{modeling} of assumptions and hypotheses, which is more conducive to accurate analyses than the formulaic application of ready-made statistics.

\section{Case Studies}
\label{sec:case-studies}

We present three case studies of applying Bayesian analysis to interpret empirical data in software engineering research.\footnote{Data analysis was done in Python using the libraries \texttt{numpy}, \texttt{scipy}, \texttt{matplotlib}, and \texttt{thinkbayes}~\cite{thinkBayes}; data and analysis scripts are available online at \url{https://bitbucket.org/caf/bayesstats-se}.}
Each case study recalls data and results in a previous publication (\emph{original data} and \emph{previous results}), and then presents novel Bayesian analyses part of the present paper's research.

\iflong
Every case study:
\begin{enumerate*}
\item presents the main data;
\item summarizes the analysis we carried out in previous research based on those data;
\item introduces additional data that provides complementary information;
\item describes a Bayesian analysis;
\item summarizes its results in terms of the case study;
\item suggests remaining aspects that deserve further investigation.
\end{enumerate*}
\fi

\subsection{Agile vs.\ Structured Development}
\label{sec:agile}

The Agile vs.\ Structured study~\cite{agile,agile-extended} (for brevity, \emph{\agile}) compares agile and heavyweight/structured software development processes based on a survey of IT companies worldwide involved in distributed and outsourced development.
Here, we target \agile's analysis of overall project success. 
\iflong
\autoref{app:agile} also discusses the project importance for customers.
\fi

\fakepar{Original data}
\agile surveyed 47 projects $P$, partitioned according to whether they followed an agile process (29 projects $P_A$) or a more heavyweight, structured process (18 projects $P_S$).\footnote{A binary classification of development processes is a simplification, but we took care~\cite[Sec.~7]{agile-extended} of limiting its impact on the validity of data.}
For each project $p \in P$, the survey's respondents assessed its \emph{outcome} $O(p)$ on a scale 1--10, where 1 denotes complete failure and 10 denotes full success.
The multiset
$O_A = \{O(p) \mid p \in P_A\}$ collects the outcome of all agile projects; 
$O_S = \{O(p) \mid p \in P_S\}$ the outcome of all structured projects; 
and $O = O_A \cup O_S$ the outcome of all projects.

\fakepar{Previous results}
\agile compared $O_A$ to $O_S$ using a $U$ test---a frequentist test applied to the null hypothesis that there is no significant difference between projects carried out using agile or using structured processes; 
the test's $p$-value was quite \emph{large}: $p = 0.571$.
Indeed, pretty much all the analyses of \agile---including many aspects other than success---failed to reject the null hypotheses that projects following agile processes and projects following structured processes behave significantly differently.
\agile concluded that there is no a priori reason to prefer agile over structured; different projects may require different approaches, and each development process can be effective in a certain domain.

\fakepar{New questions}
\agile's analysis does not comply with frequentist orthodoxy, according to which you can never ``accept'' a null hypothesis but only ``fail to reject it''.
From a Bayesian point of view, a large $p$-value does not warrant the conclusion that the null hypothesis is likely to hold (\autoref{sec:classic-vs-bayes}).
Independent of this shortcoming, \agile's results go against the general practitioners' opinion\footnote{Experimental evidence is more specific and nuanced~\cite{MullerTichy01,NawrockiETAL02,HulkkoAbrahamsson05,BhatNagappan06,BegelNagappan08}.} that agile processes are more effective than traditional, structured ones.
Can we validate \agile's analysis in a more general context with data coming from other sources?
\rquestion{
\rqu{2}:
What is the \emph{typical} impact of adopting agile rather than structured processes on the overall outcome of software development projects?
}

\fakepar{Additional data (Ambysoft study)}
As additional data on software project outcome we consider Ambysoft's IT Project Success Rates Survey~\cite{itproj} (for brevity, \emph{\itproj}).
The data from \itproj are comparable to those from \agile as they both consist of the results from surveys of a substantial number of IT professionals, explicitly classify projects into agile and structured, and target overlapping aspects.
Specifically, \itproj collected data about the ``success'' of ``software delivery teams'', which directly relates to project \emph{outcome}.

\itproj's data is organized a bit differently than \agile's.
The 173 survey respondents $R$ assessed project outcome in four categories of development processes: ad-hoc $H$, agile $A$, traditional $T$, iterative $I$, and lean $L$.
For each category $c \in \{A, H, I, L, T\}$, each respondent $r \in R$ estimated percentages $p_c^0(r)$, $p_c^1(r)$, and $p_c^2(r)$ of projects in category $c$ that were failures ($p_c^0$), challenges ($p_c^1$), and successful ($p_c^2$).\iflong\footnote{We discarded the $p_c$s of respondents who declared no experience in projects of category $c$.}\fi

\fakepar{Bayesian analysis}
To make the data in \agile and in \itproj quantitatively comparable, we adjust scales and formats, and match categories of processes.
In \agile, we introduce primed versions $O'_A, O'_S, O'$ of $O_A, O_S, O$ by uniformly rescaling the data in the unprimed sets (ranging over $[1..10]$) over the range $[0..2]$ used in \itproj's data.
Then, $\overline{d}_A(k) = \left| \{ o \in O'_A \mid o = k\}\right|$ is the number of projects in $A$ with outcome $0 \leq k \leq 2$; $\overline{d}_S(k)$ and $\overline{d}(k)$ are defined similarly for structured and for all projects.

In \itproj, for every non-empty subset $C \subseteq \{A, H, I, L, T\}$, we define distribution $\outcome_C$ over values in $[0..2]$ as follows.
If $C$ is a singleton set $\{c\}$ with $c \in \{A, H, I, L, T\}$, $\outcome_c[k]$ is the probability $\sum_{r \in R} p_c^k/|R|$ that a project in category $c$ has outcome $k$, obtained by averaging all responses.
If $C$ is a non-empty subset of $\{A, H, I, L, T\}$, $\outcome_C[k]$ is the weighted average $\outcome_C[k] = \sum_{c \in C}\outcome_{c}[k]/|C|$.
We refer to $\outcome_C$ as \emph{outcome distribution} for projects of categories $C$.
\autoref{tab:agile-distros} shows the outcome distributions for nine subsets $C$ of process categories. 

\begin{table}[ht]
\centering
\setlength{\tabcolsep}{3pt}
\begin{tabular}{c|*{9}{r}}
\hline
& \multicolumn{1}{c}{$A$}
& \multicolumn{1}{c}{$AIL$}
& \multicolumn{1}{c}{$AILT$}
& \multicolumn{1}{c}{$AIT$}
& \multicolumn{1}{c}{$AL$} 
& \multicolumn{1}{c}{$ALT$}
& \multicolumn{1}{c}{$AT$} 
& \multicolumn{1}{c}{$IT$}
& \multicolumn{1}{c}{$T$}
\\
\hline
$\outcome_{C}[{0}]$  &  7{\scriptsize\,\%}  &  8{\scriptsize\,\%}  &  10{\scriptsize\,\%}  &  11{\scriptsize\,\%}  &  7{\scriptsize\,\%}  &  11{\scriptsize\,\%}  &  12{\scriptsize\,\%}  &  12{\scriptsize\,\%}  &  18{\scriptsize\,\%} \\
$\outcome_{C}[{1}]$  &  30{\scriptsize\,\%}  &  27{\scriptsize\,\%}  &  28{\scriptsize\,\%}  &  29{\scriptsize\,\%}  &  27{\scriptsize\,\%}  &  29{\scriptsize\,\%}  &  31{\scriptsize\,\%}  &  29{\scriptsize\,\%}  &  32{\scriptsize\,\%} \\
$\outcome_{C}[{2}]$  &  63{\scriptsize\,\%}  &  65{\scriptsize\,\%}  &  62{\scriptsize\,\%}  &  60{\scriptsize\,\%}  &  66{\scriptsize\,\%}  &  60{\scriptsize\,\%}  &  57{\scriptsize\,\%}  &  59{\scriptsize\,\%}  &  50{\scriptsize\,\%} \\
\hline
\end{tabular}
\caption{Data from \itproj: for each column header $C \subseteq \{A, I, L, T\}$, $\outcome_C[k]$ is the probability that the outcome of projects following processes in categories $C$ is $k \in [0..2]$.}
\label{tab:agile-distros}
\end{table}

Since we are comparing project outcomes, we need a notion of an outcome distribution being \emph{better} than another: $p$ is better than $q$, written $p > q$, iff $\mu(p) > \mu(q)$,\footnote{The mean $\mu(p)$ of $p$ is $\sum_k k\cdot{}p[k]$.} that is if $p$ leads to higher quality outcomes than $q$ on average; otherwise we write $p \leq q$.
For example, $\outcome_A > \outcome_T$ and $\outcome_A \leq \outcome_{AL}$ in \autoref{tab:agile-distros}.

If $p$ is any outcome distribution, note that the probability that $d_0$ projects have outcome 0, $d_1$ have outcome 1, and $d_2$ have outcome 2 out of a total of $n$ projects following $p$ is given by the multinomial \pmf 
\[
M(d_0, d_1, d_2; p) = \frac{(d_0 + d_1 + d_2)!}{d_0! d_1! d_2!} p[0]^{d_0} p[1]^{d_1} p[2]^{d_2}\,.
\]

The goal of Bayesian analysis is assessing whether the data from \agile supports the hypothesis ``agile leads to more successful projects'' ($h_A$) more than the hypothesis ``agile is as good as structured'' ($h_=$).
To this end, the \emph{likelihood} function $\like_C[D; h]$ should weigh the same data $D$ differently according to whether $h = h_A$ or $h = h_=$. 
Let $D = D_A \cup D_S$ be partitioned in data $D_A$ about agile projects and data $D_S$ about structured projects.
Then, $\like_C[D; h_A]$ should assign a different weight to $D_A$ with respect to $D_S$, whereas $\like_C[D; h_=]$ should assign the same weight to $D$ for all projects regardless of their kinds (agile or structured).
If we knew accurate distributions of outcome $\pbetter$ for agile projects, $\pworse$ for structured projects, and $\pall$ for all projects, we could just compute the likelihood as $\like_C[D; h_A] = M(D_A; \pbetter) \cdot M(D_S; \pworse)$ and $\like_C[D; h_=] = M(D_A; \pall) \cdot M(D_S; \pall)$.
However, getting accurate distributions is the whole point of the analysis!
Whatever the choice of fixed $\pbetter$, $\pworse$, and $\pall$ (for example, we could base it on \autoref{tab:agile-distros}'s data), the results of the analysis would hinge on the choice, and hence risk overfitting.

Bayesian analysis, however, can average out over all possible distributions in a certain family.
We use \itproj's data only to provide a \emph{baseline} distribution $\outcome_C$. To test whether agile projects are better than structured, we assign to the former all outcome distributions that are better than the baseline (first product term in \eqref{eq:like-agile-ha}), and to the latter all outcome distributions that are worse or as good as the baseline (second term in \eqref{eq:like-agile-ha}):
\begin{equation}
\like_C[D; h_A] = 
\Big(\sum_{p > \outcome_C} w_p\cdot{}M(D_A; p)\Big) 
  \cdot 
\Big(\sum_{p \leq \outcome_C} w_p\cdot{}M(D_S; p)\Big) \,.
\label{eq:like-agile-ha}
\end{equation}
The $w_p$s are weights giving the prior probability of each distribution, which we define below.
The likelihood for hypothesis $h_=$ is similar but sums over all outcome distributions,\footnote{For simplicity, we discretize all distributions into a finite set.} modeling the hypothesis that a project's outcome is independent of the development process:
\begin{equation}
\like_C[D; h_=] = 
\Big(\sum_{p} w_p\cdot{}M(D_A; p)\Big) 
  \cdot 
\Big(\sum_{p} w_p\cdot{}M(D_S; p)\Big) \,.
\label{eq:like-agile-h0}
\end{equation}
\autoref{tab:agile-bayes-factors-summary} shows the Bayes factors $K_C(\overline{D}) = \like_C[\overline{D}; h_A] / \like_C[\overline{D}; h_=]$ for the data $\overline{D} = (\overline{d}_A(0), \overline{d}_A(1), \overline{d}_A(2)) \cup (\overline{d}_S(0), \overline{d}_S(1), \overline{d}_S(2))$ from \agile, for every distribution $C$ in \autoref{tab:agile-distros} as baseline.
Each row uses different weights $w_p$s in \eqref{eq:like-agile-ha} and \eqref{eq:like-agile-h0}: \emph{uniform} weighs all distributions equally; \emph{triangle} decreases the weigh linearly with the difference $\delta$ between $\mu(p)$ and the baseline $\mu(\outcome_C)$; \emph{power} decreases it like $(1+\delta)^{-1}$; and \emph{exp} like $\exp(-\delta)$.

\begin{table}[ht]
\centering
\setlength{\tabcolsep}{3pt}
\begin{tabular}{l|rrrrrrrrr}
\hline
& \multicolumn{1}{c}{$A$}
& \multicolumn{1}{c}{$AIL$}
& \multicolumn{1}{c}{$AILT$}
& \multicolumn{1}{c}{$AIT$}
& \multicolumn{1}{c}{$AL$} 
& \multicolumn{1}{c}{$ALT$}
& \multicolumn{1}{c}{$AT$} 
& \multicolumn{1}{c}{$IT$}
& \multicolumn{1}{c}{$T$}
\\
\hline
uniform  &  0.25  &  0.26  &  0.17  &  0.14  &  0.29  &  0.12  &  0.08  &  0.10  &  0.01 \\
triangle  &  0.25  &  0.26  &  0.17  &  0.14  &  0.29  &  0.13  &  0.08  &  0.10  &  0.02 \\
power  &  0.25  &  0.26  &  0.17  &  0.14  &  0.29  &  0.13  &  0.09  &  0.11  &  0.02 \\
exp  &  0.25  &  0.26  &  0.19  &  0.16  &  0.29  &  0.15  &  0.10  &  0.12  &  0.02 \\
\hline
\end{tabular}
\caption{Bayes factors $K_C(\overline{D})$ estimating whether the data $\overline{D}$ supports hypothesis $h_A$ (agile leads to more successful projects) more than hypothesis $h_=$ (agile is no more successful), for different baseline project outcome distributions.} 
\label{tab:agile-bayes-factors-summary}
\end{table}

\fakepar{New results}
Regardless of the choice of weights and baseline distribution, the Bayes factors in \autoref{tab:agile-bayes-factors-summary} are not significant (see \autoref{sec:classic-vs-bayes}); on the contrary, factors less than one suggest that the data supports hypothesis $h_=$ more than $h_A$.
We do not report variants of this analysis, where we rescaled the data in $O$ differently\iflong (to account for the fact that the values in $O$ do not span the entire available range $[1..10]$)\fi; 
in all cases factors do not significantly change.
Thus, Bayesian analysis confirms the results of~\cite{agile} with a stronger degree of confidence.

\takehome{\ans{2}: 
Software projects developed following an agile process do not have consistently better outcomes than projects developed following a structured process.
}

\iflong
In passing, we also largely agree with \cite{ambler-dobbs}'s conclusions that the claims of a ``software crisis'' are not supported by the evidence that software projects seem to be successful to a large degree.
\fi

\fakepar{Further analyses}
Since the Bayesian analysis confirms \agile's results, further improvements should look into whether the data can be made more rigorous.
A recurring threat follows from the observation that different IT professionals may have different views of what an ``agile process'' is.
The data in \itproj, which distinguish between categories such as ``agile'' and ``lean'' that would be natural to lump together, suggests that a sharp classification may be hard to obtain. 
Future work could collect data by inspecting individual processes to ensure that a uniform classification criterion is applied.
Note, however, that results are unlikely to change dramatically for the aspects that we analyzed: respondents already tended to give high ranks to agile projects, but this was not enough to show a significant overall difference, indicating that there are probably factors as or more important than the development process that determine a project's success.

\subsection{Programming Languages Performance Comparison}
\label{sec:rosetta}

The Rosetta code study~\cite{rosetta} (for brevity, \emph{\rosetta}) compares eight programming languages for features such as conciseness and performance, based on experiments with a curated selection of programs from the Rosetta Code repository~\cite{rosettacode}.
Here, we target \rosetta's running time performance analysis.
\iflong
\autoref{app:rosetta} also discusses the analysis of memory usage.
\fi

\fakepar{Original data}
For each language $\ell$ among C, C\#, F\#, Go, Haskell, Java, Python, and Ruby, \rosetta's \emph{performance} experiments involved a set $T(\ell)$ of programming \emph{task}, such as sorting algorithms, combinatorial puzzles, and NP-complete problems.
For each task $t$, $S(\ell, t)$ denotes the running time of the \emph{best} (that is, the fastest) implementation in language $\ell$ among those available in Rosetta Code that ran without errors or timeout on the same predefined input; $\vec{S}(\ell)$ is the set of all running time measures $S(\ell, t)$, for $t \in T(\ell)$.
For each \emph{pair} $\ell_1, \ell_2$ of languages, the set $\vec{S}(\ell_1,\ell_2)$ includes all elements 
$
S(\ell_1, \ell_2, t) \ =\ \rho(S(\ell_1, t), S(\ell_2, t)),\text{ for }t \in T(\ell_1) \cap T(\ell_2),
$
where 
\begin{equation}
\label{eq:ratio-def}
\rho(a, b)\ =\ \sgn(a - b) \frac{\max(a, b)}{\min(a, b)},
\end{equation}
and $\sgn(z) = 1$ for $z > 0$ and $\sgn(z) = -1$ for $z \leq 0$.
Note that $|S(\ell_1, \ell_2, t)| \geq 1$; thus, $S(\ell_1, \ell_2, t)$ represents the \emph{speedup} of one language over the other in task $t$: a positive value indicates that language $\ell_2$ was $|S(\ell_1, \ell_2, t)|$ times faster than language $\ell_1$ on task $t$; a negative value indicates that language $\ell_1$ was $|S(\ell_1, \ell_2, t)|$ times faster than $\ell_2$.

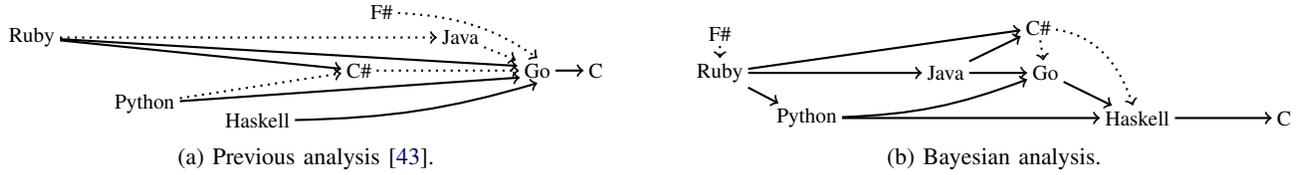
\begin{figure*}[!t]
\centering
\begin{subfigure}[t]{0.5\textwidth}
\centering
  \begin{tikzpicture}[
  lang/.style={draw=none,font=\footnotesize,inner sep=1pt,outer sep=1pt},
  align=center, xscale=0.3, yscale=0.22
  ]
\node [lang] (C) at (1,3) {C};
\node [lang] (C-sharp) at (-9.5,3) {C\#};
\node [lang] (F-Sharp) at (-8.5,6.5) {F\#};
\node [lang] (Go) at (-1.6,3) {Go};
\node [lang] (Haskell) at (-14,0) {Haskell};
\node [lang] (Java) at (-5,5) {Java};
\node [lang] (Python) at (-19,1) {Python};
\node [lang] (Ruby) at (-24,5) {Ruby};
\draw [->,thick](Go) edge (C);
\draw [->,dotted,thick](C-sharp) edge ($(Go.west)+(-0pt,-0pt)$);
\draw [->,dotted,thick](Python) edge (C-sharp);
\draw [->,thick](Ruby) edge (C-sharp);
\draw [->,dotted,thick,bend left=20](F-Sharp) edge (Go.north);
\draw [->,thick,bend right=10](Haskell) edge (Go.south);
\draw [->,dotted,thick](Java) edge ($(Go.west)+(-4pt,18pt)$);
\draw [->,thick](Python) edge ($(Go.west)+(1pt,-12pt)$);
\draw [->,thick](Ruby) edge ($(Go.west)+(-2pt,8pt)$);
\draw [->,dotted,thick](Ruby) edge (Java);
  \end{tikzpicture}
\caption{Previous analysis~\cite{rosetta}.}
\label{fig:OLD-network_running}
\end{subfigure}%
~%
\begin{subfigure}[t]{0.5\textwidth}
\centering
  \begin{tikzpicture}[
  lang/.style={draw=none,font=\footnotesize,inner sep=1pt,outer sep=1pt},
  align=center, xscale=0.3, yscale=0.4
  ]
\node [lang] (C) at (13,-1.5) {C};
\node [lang] (C sharp) at (2.10706346328997,1.5) {C\#};
\node [lang] (F sharp) at (-12,1.3) {F\#};
\node [lang] (Go) at (2.42161483433402,0) {Go};
\node [lang] (Haskell) at (6.5,-1.5) {Haskell};
\node [lang] (Java) at (-2,0) {Java};
\node [lang] (Python) at (-8.15371079415328,-1.5) {Python};
\node [lang] (Ruby) at (-12,0) {Ruby};
\draw [->,thick,dotted,bend left](C sharp) edge (Haskell);
\draw [->,thick](Haskell) edge (C);
\draw [->,thick,dotted](C sharp) edge (Go);
\draw [->,thick](Java) edge (C sharp);
\draw [->,thick](Ruby) edge (C sharp);
\draw [->,thick,dotted](F sharp) edge (Ruby);
\draw [->,thick](Go) edge (Haskell);
\draw [->,thick](Java) edge (Go);
\draw [->,thick,bend right=7](Python) edge (Go);
\draw [->,thick](Python) edge (Haskell);
\draw [->,thick](Ruby) edge (Java);
\draw [->,thick](Ruby) edge (Python);
  \end{tikzpicture}
\caption{Bayesian analysis.}
\label{fig:network_running}
\end{subfigure}
\label{fig:both_network_running}
\caption{Comparison of running time: qualitative summaries.}
\end{figure*}

\fakepar{Previous results}
For each pair $\ell_1, \ell_2$ of languages, \rosetta compared $\vec{S}(\ell_1)$ to $\vec{S}(\ell_2)$ using:
\begin{enumerate*}
\item a Wilcoxon signed-rank test---a frequentist hypothesis test giving a $p$-value;

\item Cohen's $d$ effect size---a standardized mean difference between $\vec{S}(\ell_1)$ and $\vec{S}(\ell_2)$;

\item a signed ratio $R$---an unstandardized mean speedup between $\vec{S}(\ell_1)$ and $\vec{S}(\ell_2)$ (similar to $\vec{S}(\ell_1, \ell_2)$ but using the median running time across all tasks).
\end{enumerate*}
A \emph{language relationship graph} summarized all comparisons: nodes are languages; the horizontal distance between two nodes $\ell_1, \ell_2$ is roughly proportional to the absolute value of $R$ for those nodes; an arrow from $\ell_1$ to $\ell_2$ denotes that the corresponding $p$-value is small ($p < 0.05$), the effect size $d$ is not negligible ($d \geq 0.05$), and $\ell_2$ is faster on average ($R > 0$); if the $p$-value is $0.01 \leq p < 0.05$ the arrow is dotted to indicate lower confidence.
\autoref{fig:OLD-network_running}, copied from~\cite{rosetta}, shows the graph.\footnote{Unlike~\cite{rosetta}, we do not consider arrow thickness to indicate effect size.}

\fakepar{New questions}
\rosetta's reliance on scalar statistics such effect sizes in addition to $p$-values mitigates threats to the validity of its results; however, several language comparisons remain inconclusive.
For example, it is somewhat surprising that \rosetta could not ascertain that a compiled highly-optimized language like Haskell is generally faster than the dynamic scripting languages Python and Ruby.
We would also like to track down the impact of experimental choices that depended on factors \rosetta could not fully control for, such as which implementations were available in Rosetta Code.
\rquestion{
\rqu{1}: 
Which programming languages have better running time performance, after taking into account the potential sources of bias in \rosetta's experimental data~\cite{rosetta}?
}

\fakepar{Additional data (benchmarks)}
As additional data on performance and memory usage we consider the Computer Language Benchmarks Game~\cite{benchmarks} (for brevity, \emph{\bench}).
The data from \bench are comparable to those from \rosetta as they both consist of curated selections of collectively written solutions to well-defined programming tasks running on the same input and refined over a significant stretch of time;\iflong\footnote{Some details of the performance measures are also similar, such as the choice of including the Java VM startup time in the running time measures.}\fi{} on the other hand, \bench was developed independently of \rosetta, which makes it a complementary source of data.

For each language $\ell$, \bench's \emph{performance} experiments determine a set $\bvec{S}(\ell)$ with elements $\b{S}(\ell, t, n, v)$, for $t$ ranging over the set $\b{T}(\ell)$ of \bench's tasks, $n$ ranging over the set $\b{N}(\ell,t)$ of input sizes of task $t$ in $\ell$, and $v$ ranging over the set $\b{V}(\ell,t)$ of different implementations of the same task $t$ in $\ell$.
\bench's tasks include numerical algorithms, regular expression matching, and algorithms on trees.
\bench's performance data include experiments with different solutions for the same task and inputs of different sizes; we avail this to model the possible variability in performance measurements.
For each \emph{pair} $\ell_1, \ell_2$ of languages, the set $\bvec{S}(\ell_1, \ell_2)$ includes all elements 
\[
\b{S}(\ell_1,\ell_2, t) \ =\ 
\rho\Big(
\min_{v \in \b{V}(\ell_1,t)} \b{S}(\ell_1, t,\overline{m},v),
\min_{v \in \b{V}(\ell_2,t)} \b{S}(\ell_2, t,\overline{m},v)
\Big),
\]
for $t \in \b{T}(\ell_1) \cap \b{T}(\ell_2)$ and $\overline{m} = \max(\b{N}(\ell_1, t) \cap \b{N}(\ell_2, t))$; 
that is, $\b{S}(\ell_1,\ell_2, t)$ is the speedup ratio \eqref{eq:ratio-def} of the fastest solution in $\ell_1$ over the fastest solution in $\ell_2$ for the same task $t$ and running over the largest input that both languages can handle.
Thus, $\bvec{S}(\ell_1, \ell_2)$ is directly comparable to $\vec{S}(\ell_1, \ell_2)$ as the similar notation suggests.

We also define the set $\bvec{S}_{\Delta}(\ell_1, \ell_2, t)$ of all values $\rho(\bvec{S}(\ell_1, t, n, v_1), \bvec{S}(\ell_1, t, n, v_2)) - \b{S}(\ell_1, \ell_2, t)$, for $n$ ranging over $\b{N}(\ell_1, t) \cap \b{N}(\ell_2, t)$, $v_1$ ranging over $\b{V}(\ell_1, t)$, and $v_2$ ranging over $\b{V}(\ell_2, t)$; intuitively, $\bvec{S}_{\Delta}(\ell_1, \ell_2, t)$ is the distribution of all differences in speedup measurements for task $t$ between any two programs (on input of any size) and the two fastest programs (on the largest input).
$\bvec{S}_{\Delta}(\ell_1, \ell_2, t)$ gives an idea of the variability in speedup ratios that may result from inputs or programs other than those that turned out to be the fastest. 

\fakepar{Bayesian analysis}
For every pair $\ell_1, \ell_2$ of languages, the \emph{prior} distribution $\prior^S(\ell_1, \ell_2)$ gives the probability $\prior^S(\ell_1, \ell_2)[r]$ of observing a program in $\ell_1$ and a program in $\ell_2$---solving the same problem and input---respectively running for $t_1$ and $t_2$ time units such that $\rho(t_1, t_2) = r$.
Informally, the prior models the initial expectations on the performance difference between languages---which one will be faster and how much.
We base our initial expectations on the results of \bench; hence, $\prior^S(\ell_1, \ell_2)$ follows the distribution of $\bvec{S}(\ell_1, \ell_2)$.
Precisely, $\bvec{S}(\ell_1, \ell_2)$ is based on a finite number of discrete observations, and hence it excludes values that are perfectly acceptable but did not happen to occur in the experiments.
But if, say, $\ell_2$ is twice as fast as $\ell_1$ in an experiment and three times as fast in another experiment, we expect speedup values between $2$ and $3$ to be possible even if they were not observed in any performed experiment.
Thus, $\prior^S(\ell_1, \ell_2)$ is the kernel density estimation (KDE~\cite{kde-book} using a normal kernel function\iflong\footnote{We used Python's \texttt{scipy.stats.gaussian\_kde} function.}\fi{}) of $\bvec{S}(\ell_1, \ell_2)$; furthermore, we rework the smooth distribution obtained by KDE to exclude values in the interval $(-1, 1]$ since these are impossible given the definition of $\rho$ in \eqref{eq:ratio-def}.

The \emph{likelihood} $\like^S(\ell_1, \ell_2)[d; h]$ expresses how likely observing a speedup $d$ is, under the hypothesis that the actual speedup is $h$.
We base it on \bench's extended experiments following this argumentation.
The outcome of performance experiments also depends on some parameters, such as the input size and specific implementation choices, that are somewhat accidental; 
for example, \rosetta's experiments used inputs of significant size, manually selected; these choices seem reasonable, but we cannot exclude that, if input sizes had been chosen differently, the performance results would have been quantitatively different.
In order to assess this experimental uncertainty due to effects that cannot be entirely controlled, we base the likelihood on the values $\bvec{S}_{\Delta}(\ell_1, \ell_2, t)$, which span the differences between the reported data $\bvec{S}(\ell_1, \ell_2)$ and the same metric for different choices of input size or program variant.
Similarly to what we did for the prior, we smooth the distribution of values $\bigcup_t \bvec{S}_{\Delta}(\ell_1, \ell_2, t)$ using KDE\footnote{We take the union over all tasks in \bench because they differ from \rosetta's.}, which yields a probability density function $\Delta(\ell_1, \ell_2)$.
Then, the likelihood $\like^S(\ell_1, \ell_2)[d; h] \propto \Delta(\ell_1, \ell_2)[d - h]$ is a value proportional to the probability of observing the difference $d - h$ of speedups. 

\begin{figure*}[!t]
\centering
\begin{subfigure}[t]{0.333\textwidth}
\centering
\includegraphics[width=\textwidth]{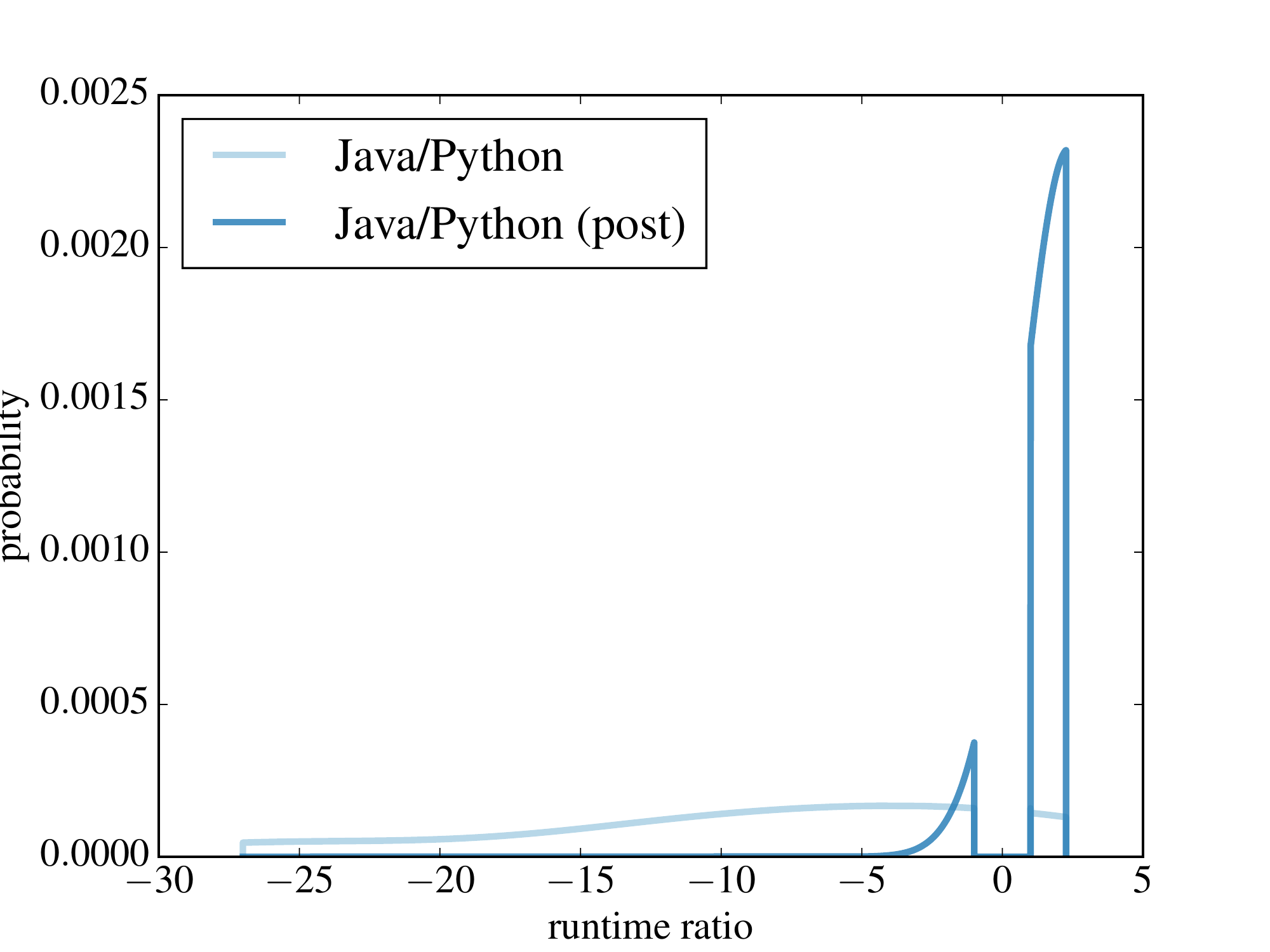}
\end{subfigure}%
~%
\begin{subfigure}[t]{0.333\textwidth}
\centering
\includegraphics[width=\textwidth]{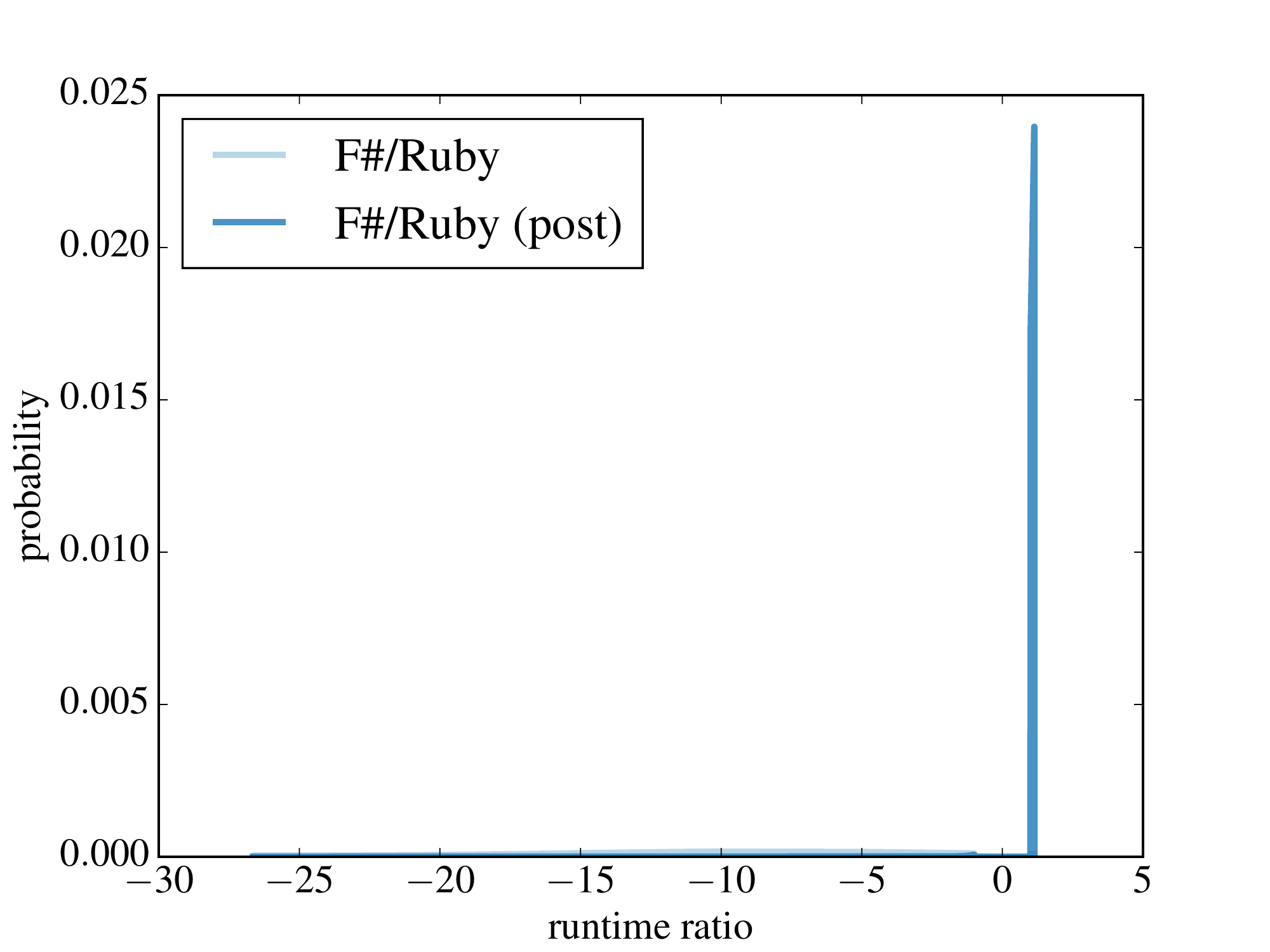}
\end{subfigure}
\caption{Posterior distributions $\post^S_{\vec{S}}$ of running time ratios of Java vs.\ Python (left) and F\# vs.\ Ruby (right).}
\label{fig:posteriors-performance}
\end{figure*}

The posterior distribution $\post^S_{\vec{S}}(\ell_1, \ell_2)$ is obtained by updating the prior (\autoref{sec:classic-vs-bayes}) with \rosetta's data $\vec{S}$; $\post^S_{\vec{S}}(\ell_1, \ell_2)[r]$ is the probability that the speedup of $\ell_1$ over $\ell_2$ is $r$.
\autoref{tab:running} summarizes the posteriors using two statistics: CI is the 95\% credible interval\footnote{Credible intervals are Bayesian analogues of confidence intervals.} (that is, there is a 95\% chance that the real speedup falls in the interval), and $m$ is the median.\iflong\footnote{The means are generally close to the medians.}\fi{} 
Credible intervals that include $0$ may indicate an inconclusive comparison (one or the other language may be faster)
One advantage of Bayesian analysis is that it provides \emph{distributions} (rather than just scalar summaries), so that we can sort out borderline cases by visually inspecting them.
For example, \autoref{fig:posteriors-performance} suggests that the Java vs.\ Python comparison is indeed inconclusive (there's significant probability on both sides of the origin), whereas the F\# vs.\ Ruby comparison has a very sharp peak next to $1$, which suggests that Ruby was consistently faster than F\#---albeit not \emph{much} faster.
We summarize the results of the posteriors' analysis in the language relationship graph in \autoref{fig:network_running}.
It conveys the same general information as the graph in \autoref{fig:OLD-network_running} from~\cite{rosetta}, but it is based on Bayesian analysis; now, a dotted arrow indicates a speedup relationship that is weak or borderline but still likely to hold (such as F\# vs.\ Ruby).

\begin{table}[ht]
\centering
\tiny
\setlength{\tabcolsep}{1pt}
\renewcommand{\notsig}[1]{#1}
\renewcommand{\weaksig}[1]{#1}
\renewcommand{\strongsig}[1]{#1}
\begin{tabular}{cc|rrrrrrr}
  \hline
\textsc{language} &  & \multicolumn{1}{c}{C} & \multicolumn{1}{c}{C\#} & \multicolumn{1}{c}{F\#} & \multicolumn{1}{c}{Go} & \multicolumn{1}{c}{Haskell} & \multicolumn{1}{c}{Java} & \multicolumn{1}{c}{Python} \\ 
  \hline
C\# & CI & \strongsig{(-10.1, -8.7)} &  &  &  &  &  &  \\ 
   & $m$ & -9.22 &  &  &  &  &  &  \\ 
   \hline
F\# & CI & \strongsig{(-76.0, -64.5)} & \strongsig{(-8.9, -4.6)} &  &  &  &  &  \\ 
   & $m$ & -72.61 & -5.29 &  &  &  &  &  \\ 
   \hline
Go & CI & \strongsig{(-2.3, -1.3)} & \weaksig{(1.0, 2.5)} & \strongsig{(16.9, 20.6)} &  &  &  &  \\ 
   & $m$ & -1.67 & 1.15 & 18.21 &  &  &  &  \\ 
   \hline
Haskell & CI & \strongsig{(-5.9, -5.7)} & \strongsig{(1.2, 1.7)} & \strongsig{(3.2, 15.5)} & \strongsig{(2.4, 2.5)} &  &  &  \\ 
   & $m$ & -5.76 & 1.23 & 6.77 & 2.49 &  &  &  \\ 
   \hline
Java & CI & \strongsig{(-3.2, -3.1)} & \strongsig{(-2.0, -1.3)} & \strongsig{(5.7, 7.5)} & \strongsig{(-8.5, -7.6)} & \strongsig{(-8.6, -8.3)} &  &  \\ 
   & $m$ & -3.18 & -1.77 & 6.94 & -8.02 & -8.63 &  &  \\ 
   \hline
Python & CI & \strongsig{(-54.2, -32.3)} & \notsig{(-1.4, 1.8)} & \strongsig{(2.1, 12.7)} & \strongsig{(-27.2, -17.7)} & \weaksig{(-5.0, -1.2)} & \notsig{(-2.1, 2.2)} &  \\ 
   & $m$ & -52.47 & 1.3 & 7.93 & -23.01 & -1.82 & 1.76 &  \\ 
   \hline
Ruby & CI & \strongsig{(-124.0, -90.9)} & \strongsig{(-21.1, -11.6)} & \notsig{(1.0, 1.1)} & \strongsig{(-142.2, -36.9)} & \strongsig{(-22.0, -19.9)} & \strongsig{(-17.0, -8.6)} & \strongsig{(-19.7, -14.0)} \\ 
   & $m$ & -100.69 & -16.96 & 1.05 & -141.92 & -21.85 & -15.32 & -15.03 \\ 
   \hline
\end{tabular}
\caption{Comparison of running time: each cell in column $\ell_1$ and row $\ell_2$ reports the 95\% credible interval CI and the median $m$ of the posterior distribution $\post^S_{\vec{S}}(\ell_1, \ell_2)$.} 
\label{tab:running}
\end{table}

\fakepar{New results}
Compared with \rosetta's frequentist analysis~\cite{rosetta}, the overall picture emerging from Bayesian analysis is richer and somewhat more nuanced.
C remains the king of speed, but Go cannot claim to stand out as lone runner up: Haskell is faster than Go on average (it was slower in the previous analysis), even though the performance advantage of C over Haskell is still greater than its advantage over Go.
On the other hand, several comparisons that were surprisingly inconclusive in \rosetta are now more clearly defined.
Haskell emerges as faster than the scripting languages (Python and Ruby) and than the bytecode object-oriented languages (C\# and Java).
In the opposite direction, F\# has shown a generally poor performance---in particular, quite slower than C\# even if they both run on the same .NET platform.
These differences indicate that a few results of \rosetta hinged on contingent experimental details; Bayesian analysis has lessened the bias by incorporating an independent data source.

\takehome{\ans{1}: C is the king of performance. Go and Haskell (which compile to native) are the runner-ups. 
Object-oriented languages (C\#, Java) retain a competitive performance on several tasks even if they compile to bytecode.
Interpreted scripting languages (Python, Ruby) tend to be the slowest. }

\fakepar{Further analyses}
Since Bayesian analysis relies on the data from \bench, further analysis could try different sources for the prior and likelihood distributions, in order to understand the sensitivity of the analysis on the particular choice that was done.
\bench, however, was chosen because it is the only data we could find that is publicly available, described in detail, and sufficiently similar to \rosetta to be comparable to it; thus getting more data may require to perform new experiments.

Another natural continuation of this work could collect additional data specifically for the comparisons where significant uncertainty remains.
More data about C is probably redundant as its role as performance king is largely undisputed.
In contrast, F\#'s data are unsatisfactory because they often show a large variability and disappointing results for a language that compiles to the same .NET platform as C\#; more data would help explain whether F\#'s performance gap is intrinsic, or mainly due to a less mature language support. 

\subsection{Testing with Specifications}
\label{sec:testing}

The Testing with Strong Specifications paper~\cite{testing} (for brevity, \emph{\testing}) assesses the effectiveness of random testing using as oracles functional specifications in the form of assertions embedded in the code (contracts).

\fakepar{Original data}
\testing's experiments targeted the EiffelBase library, comprising 21 classes implementing data structures---such as arrays, lists, hash tables, and trees---and iterators.
\testing tested EiffelBase twice using the same random tester AutoTest: once using the simple specifications that come with EiffelBase's code, and once using stronger specifications written as part of \testing's research.
For each class $C_k$, $k = 1, \ldots, 21$, testing using simple specifications detected $\original_k$ bugs, whereas testing using strong specifications detected $\mbc_k$ bugs.
These are actual specification violations that expose genuinely incorrect behavior.
$\voriginal = t_1, \ldots, t_{21}$ and $\vmbc = \mbc_1, \ldots, \mbc_{21}$ are the sets of all bugs found using simple and using strong specifications.

\fakepar{Previous results}
\testing compared $\voriginal$ to $\vmbc$ using a Wilcoxon signed-rank test---a frequentist hypothesis test giving a $p$-value $0.006$, which lead to rejecting the null hypothesis that using simple specifications and using strong specifications makes no difference in testing effectiveness.
Also based on other data---such as the effort spent writing strong specifications---\testing argued that strong specifications\iflong{} bring significant benefits to random testing and\fi{} achieve an interesting trade-off between effort and bug-detection effectiveness.

\fakepar{New questions}
\testing's analysis is quite convincing as it stands, because it is based on analyses other than hypothesis testing; rather than confirming its results using Bayesian statistics, we extend its analysis into a different direction: studying the distribution of bugs in classes.

\rquestion{
\rqu{3}:
What is the distribution of bugs in classes? Does it satisfy the Pareto principle: ``80\% of the bugs are located in only 20\% of the classes'', or, conversely, ``80\% of the classes are affected by only 20\% of the bugs''?
}

\fakepar{Additional data}
Zhang suggested~\cite{weibull-testing} that bug distributions in modules follow a Weibull---a continuous distributions with positive parameters $\alpha$ and $\beta$, \pdf 
$
w_{\alpha, \beta}[x] \ =\ 
\left(\beta/\alpha\right)
\left(
x/\alpha
\right)^{\beta - 1}
\exp(-\left(x/\alpha\right)^\beta)
$
and \cdf\footnote{A cumulative distribution function (\cdf) $X[x]$ gives $\prob[X \leq x]$.}
\begin{equation}
W_{\alpha, \beta}[x] \ =\ 
1 - 
\exp\left(-\left(\frac{x}{\alpha}\right)^\beta\right)\!.
\label{eq:weibull-cdf}
\end{equation}
Saying that the bug distribution in modules follows a Weibull with \cdf \eqref{eq:weibull-cdf} means that a fraction $W_{\alpha, \beta}[x]$ of the modules has $x$ or fewer bugs; or, equivalently, that a random module has $x$ or fewer bugs with probability $W_{\alpha, \beta}[x]$.
Under these conditions, the Pareto principle would hold only for certain values of $\alpha$ and $\beta$: while $\beta$ determines the distribution's \emph{shape}, and hence qualitative properties such as the Pareto principle, $\alpha$ determines the distribution's \emph{scale}, and hence only specific quantitative properties.

\fakepar{Bayesian analysis: Pareto principle}
Classes are modules in object-oriented programs; thus, we can use Bayes theorem to estimate $\alpha$ and $\beta$ such that a Weibull with \cdf $W_{\alpha, \beta}$ fits the distribution of bugs $\vmbc$ \emph{detected using strong specifications}.
Using Bayesian analysis, we infer a \emph{multivariate} distribution $m$ of values for parameters $\alpha$ and $\beta$.
Since we have no inkling of plausible values for $\alpha$ and $\beta$, we use an uninformative uniform prior $\prior[\alpha, \beta] \propto 1$ for all $\alpha, \beta$ within a broad range.
The likelihood $\like[d; \alpha, \beta]$ reflects the probability that $d$ is drawn from a Weibull with parameters $\alpha$ and $\beta$; thus $\like[d; \alpha, \beta] \propto w_{\alpha, \beta}[d + 1]$, where we shift the \pdf by one unit to account for classes with no bugs.
By applying Bayes theorem, the joint posterior distribution is:
\[
m[\alpha, \beta] 
= \post_{\vmbc}[\alpha, \beta] 
= 
\nu \prod_{d \in \vmbc} w_{\alpha, \beta}[d + 1]\,,
\]
where $\nu$ is a normalization factor obtained by the, by now familiar, update rule (\autoref{sec:classic-vs-bayes}), using data $\vmbc$ from testing with strong specifications.
\begin{figure}
\centering
\includegraphics[width=0.8\columnwidth]{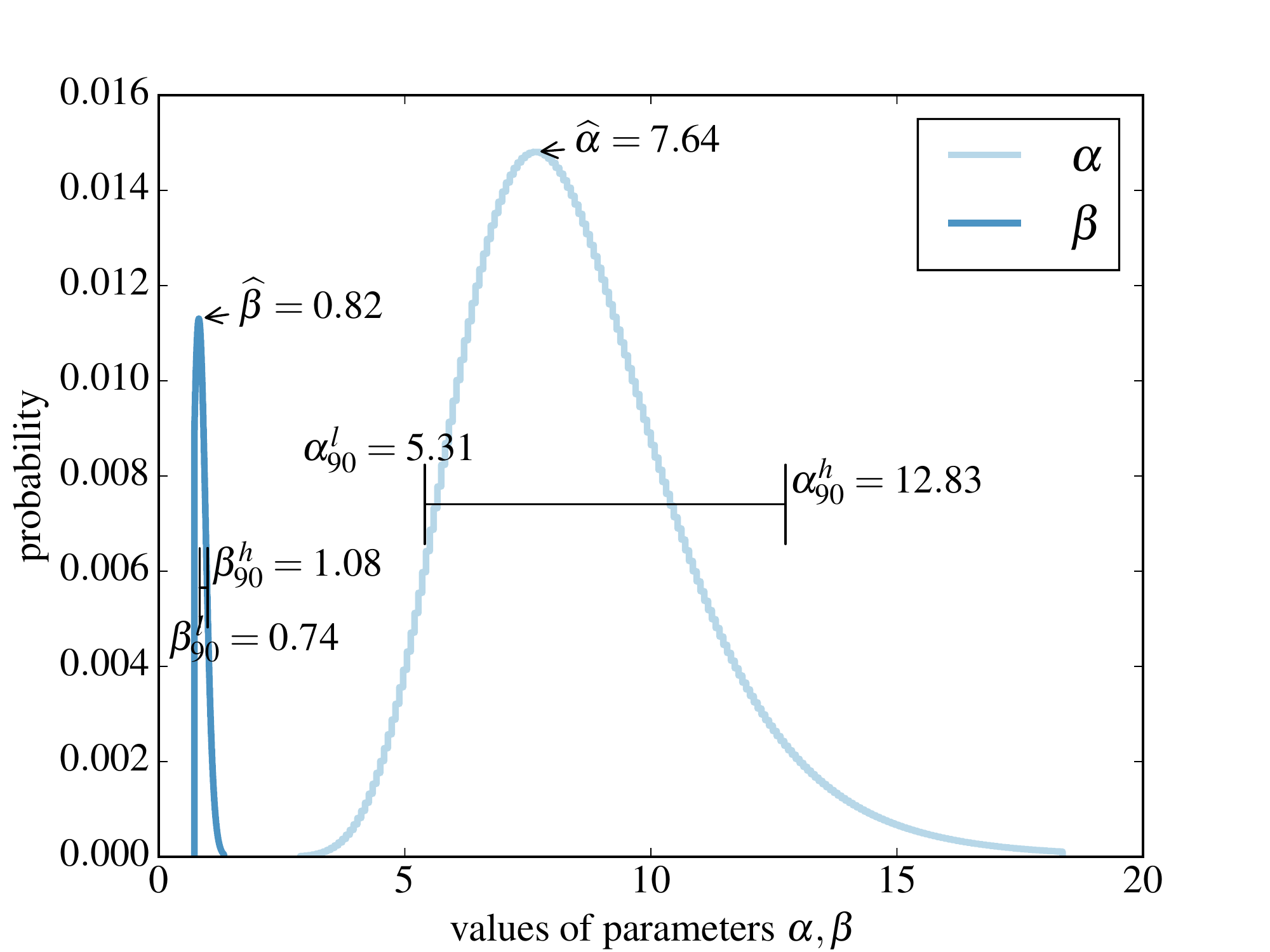}
\caption{Marginals $m[\alpha]$ and $m[\beta]$ of the posterior distribution $m[\alpha, \beta]$ of parameters $\alpha$ and $\beta$. The graph indicates maxima $\widehat{\alpha}$ and $\widehat{\beta}$ and 90\% credible intervals $(\alpha^l_{90}, \alpha^h_{90})$ and $(\beta^l_{90}, \beta^h_{90})$.}
\label{fig:testing-marginals}
\end{figure}
\autoref{fig:testing-marginals} shows $m$'s marginals $m[\alpha]$ and $m[\beta]$.\iflong\footnote{The maxima are close to means ($\mu(m[\alpha]) = 8.53$, $\mu(m[\beta]) = 0.88$) and medians ($m(m[\alpha]) = 6.86$ and $m(m[\beta]) = 0.81$).}\fi{} 
The plot indicates that there is limited uncertainty about the value of $\beta$, whereas the uncertainty about $\alpha$ is significant.
In terms of the resulting Weibull distributions, the uncertainty is mainly on the scale of the distribution (parameter $\alpha$) but not so much on its shape (parameter $\beta$).
\autoref{fig:testing-posteriors} shows this by plotting the Weibull's \cdf $W_{\alpha, \beta}$ for parameters in the 90\% credible intervals highlighted in \autoref{fig:testing-marginals}.
The picture suggests that the Pareto principle holds: the number $b$ of bugs such that $W_{\alpha, \beta}[b] = 0.8$ is 8\%, 10\%, and 13\% of the total number of possible bugs---one percentage for each choice of $\alpha, \beta$ in \autoref{fig:testing-posteriors}---which is in the ballpark of Pareto's 80--20 proportion.
The qualitative conclusions wouldn't change if we used data $\voriginal$ from testing with simple specifications.

\begin{figure}
\centering
\includegraphics[width=0.8\columnwidth]{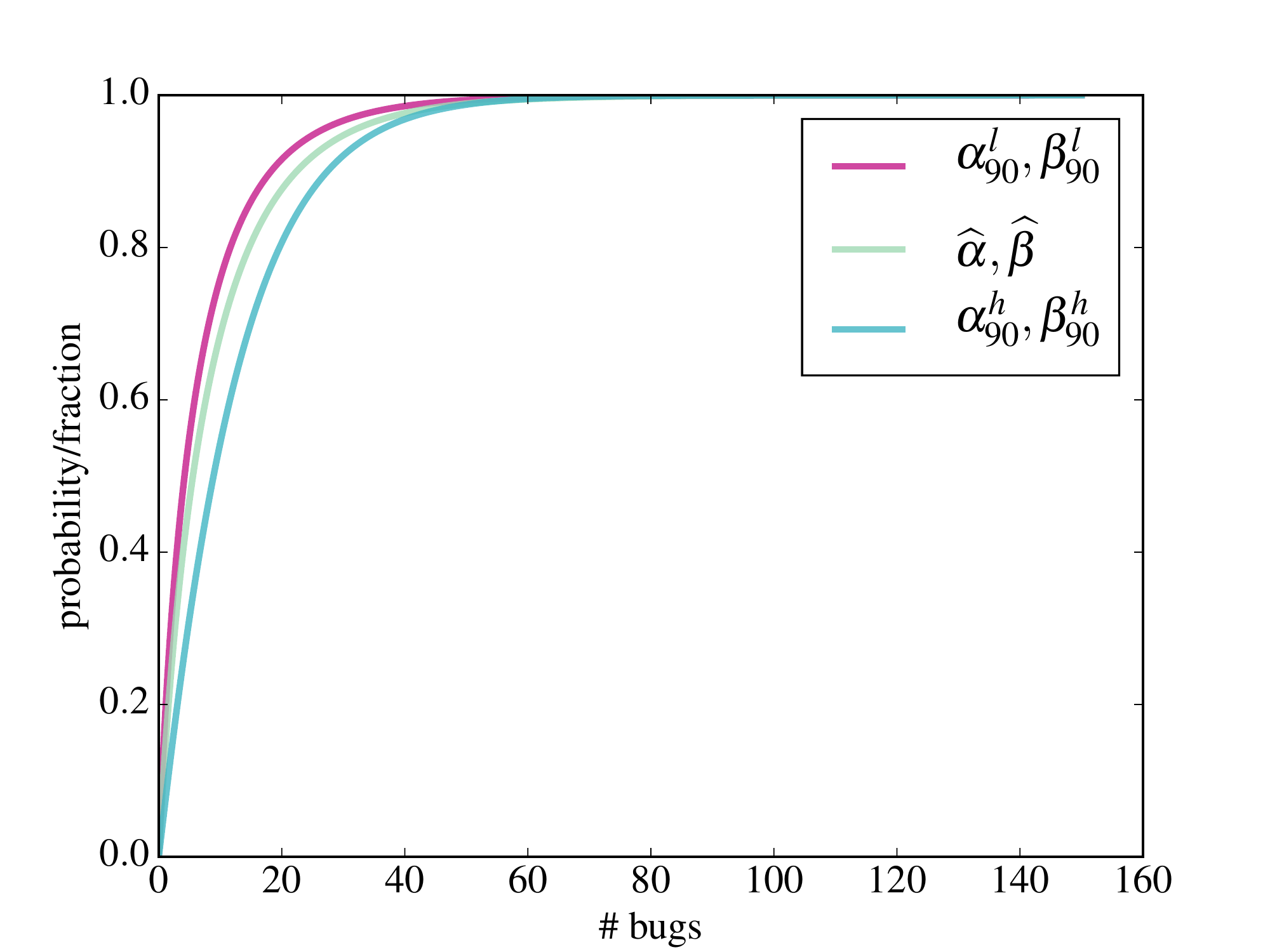}
\caption{Cumulative distribution function $W_{\alpha, \beta}$ \eqref{eq:weibull-cdf} for the different values of $\alpha$ and $\beta$ highlighted in \autoref{fig:testing-marginals}.}
\label{fig:testing-posteriors}
\end{figure}

\takehome{\ans{3-A}: 
The distribution across classes of bugs found by random testing with specifications is modeled accurately by a Weibull distribution that satisfies the Pareto principle.}

\iflong
A significant advantage of Bayesian analysis over using frequentist statistics is that we have \emph{distributions} of likely parameter values, not just pointwise estimates.
This entails that we can derive distributions of related variables.
For example, we could plot how the probability of finding a class with at most $N$ bugs--for any given $N$---varies with $\alpha$ and $\beta$.
See \autoref{app:testing} for an example of this. 
\fi

\fakepar{Bayesian analysis: total bugs}
This analysis modeled the number of bugs found by random testing; what about the \emph{total} number of bugs present in a class? Can we use Bayesian analysis to estimate it as well?

As a first step, suppose that the \emph{effectiveness} of random testing with strong specifications is $E$: if testing finds $N$ bugs---a fraction $E$ of the total---there are actually $N/E$ bugs in the class.
Similarly, let $e$ be the effectiveness of random testing with simple specifications.
Given $E$ and $e$, we can estimate the distribution $B$ of real bugs using Bayesian analysis.
The prior distribution has \pdf $\prior^b(\alpha, \beta, E)$ such that $\prior^b(\alpha, \beta, E)[x] = w_{\alpha, \beta}[x\cdot E]$, corresponding to a Weibull scaled so as to follow the expected actual bugs.
The likelihood $\like^b(e)[d; h]$ is proportional to the probability that testing with effectiveness $e$ finds $d$ bugs in a class with $h$ total bugs; thus, $\like^b(e)[d; h] \propto \mathcal{B}(h, e)[d]$, where $\mathcal{B}(h, e)$ is the binomial distribution's \pmf giving the probability of $d$ successes ($d$ bugs found) out of $h$ attempts when each attempt has probability $e$ of success.
With these prior and likelihood, the posterior distribution $\post_d^b(\alpha, \beta, e, E)[x] = B_{\alpha, \beta}(d, e, E)[x]$ gives the probability that a class has a total of $x$ bugs given that testing with effectiveness $e$ found $d$ bugs (and testing with effectiveness $E$ determined a Weibull with parameters $\alpha, \beta$).

This analysis requires knowing plausible values for $\alpha$ and $\beta$---which we can obtain from the previous analysis summarized in \autoref{fig:testing-marginals}---as well as for $e$ and $E$---which is instead the rub of the analysis.
Fortunately, we can add one layer of Bayesian inference to abstract over the unknown effectiveness values.
The uninformative prior $\prior^n_{\alpha, \beta}(d)$ is now a uniform distribution over distributions such that $\prior^n_{\alpha, \beta}(d)[e, E]$ is the probability associated with 
$B_{\alpha, \beta}(d, e, E)$ defined in the previous analysis.
The likelihood $\like^n[d; e, E]$ measures the probability that testing modeled by a distribution with parameters $e, E$ finds $d$ bugs in a class: $\like^n[d; e, E] \propto \sum_{h} \like^b(e)[d; h]\cdot B_{\alpha, \beta}(d, e, E)[h]$.
With the usual update rule, compute the posterior $\post^{\alpha, \beta}_d[e, E]$ given values for $\alpha$, $\beta$, and a number of bugs $d$ detected in some class.
Finally, $N_m^{\alpha, \beta}[n]$, which gives the probability that class $m$ has $n$ bugs, is a mixture that interpolates posteriors: 
\[
N_d^{\alpha, \beta}[n] = 
\sum_{e, E} B_{\alpha, \beta}(d, e, E)[n] \cdot \post_{d}^{\alpha, \beta}[e, E]\,,
\]
where $d$ is the number of bugs found in a class by testing with effectiveness $e$. 

\begin{table}
\centering
\tiny
\setlength{\tabcolsep}{1pt}
\begin{tabular}{l| *{21}{r}}
& $C_1$  & $C_2$  & $C_3$  & $C_4$  & $C_5$  & $C_6$  & $C_7$  & $C_8$  & $C_9$  & $C_{10}$ & $C_{11}$ & $C_{12}$ & $C_{13}$ & $C_{14}$ & $C_{15}$ & $C_{16}$ & $C_{17}$ & $C_{18}$ & $C_{19}$ & $C_{20}$ & $C_{21}$ \\
\hline
$m/|\text{M}|$  &  0.27  &  0.03  &  0.20  &  0.38  &  0.52  &  0.17  &  0.15  &  0.15  &  0.39  &  0.23  &  0.11  &  0.15  &  0.43  &  0.15  &  1.04  &  0.08  &  0.15  &  0.51  &  0.15  &  1.21  &  0.67 \\
$m(N)$  &  14  &  3  &  5  &  19  &  38  &  5  &  10  &  2  &  31  &  20  &  5  &  3  &  22  &  13  &  22  &  8  &  1  &  71  &  3  &  67  &  2 \\
$N_{90}^l$  &  5  &  0  &  2  &  18  &  20  &  2  &  2  &  0  &  2  &  2  &  2  &  0  &  22  &  0  &  4  &  4  &  0  &  25  &  0  &  49  &  0 \\
$N_{90}^h$  &  26  &  7  &  17  &  54  &  57  &  17  &  17  &  7  &  17  &  17  &  17  &  7  &  60  &  7  &  22  &  22  &  7  &  66  &  7  &  100  &  7 \\
\hline
\end{tabular}
\caption{Median per public method $m/|\text{M}|$, median $m(N)$, and 90\% credible interval $(N_{90}^l, N_{90}^h)$ of $N_{d_m}^{\na, \nb}$---estimating the total number of bugs in class $C_m$.}
\label{tab:nbugs-summary}
\end{table}

\autoref{tab:nbugs-summary} shows statistics about $N_{d_m}^{\na, \nb}$ for all 21 classes analyzed in \testing.
The parameters $\alpha = \na$ and $\beta = \nb$ are the maximum likelihood values in \autoref{fig:testing-marginals}; prior effectiveness ranges over $0.15 \leq e \leq 0.5$ for testing with simple specifications and over $0.7 \leq E \leq 0.95$ for testing with strong specifications; and $d_m = t_m$, for $m = 1, \ldots, 21$, is the number of bugs found in class $C_m$ by testing with simple specifications in \testing's experiments.
The median bugs per public method---similar to bugs per function point~\cite{Jones13}---is an indicator of bug proneness considered more robust than bugs per line of code.
According to this metric, trees (class $C_{20}$) and linked stacks (class $C_{15}$) data structures are the faultiest, while arrayed lists (class $C_2$) and linked lists (class $C_{11}$) are the least faulty.
The difference can be explained in terms of which structures are the most used in Eiffel programs: lists are widely used, and hence their implementations have been heavily tested and fixed.

\takehome{\ans{3-B}: 
The number of total bugs in a class can be estimated by Bayesian analysis from the bugs found by random testing. Classes that are less used are more error prone.}

\fakepar{Further analyses}
Using Bayesian analysis we obtained a reliable estimate of the real bugs present in a data structure library; there remains a significant margin of uncertainty, given that we abstracted over several unknown details, but the uncertainty is quantified and upheld by precise modeling choices.
Generalizing the analysis to include other testing techniques (e.g., manual testing) or, conversely, specialize it to other domains and conditions to make it more precise are natural extensions of this work.
We used a very simple model of testing effectiveness based on detection effectiveness; using more detailed models of random testing~\cite{ArcuriIB12} may provide additional insights and more accurate estimates.

\section{Threats to Validity}
\label{sec:threats}

Do Bayesian techniques help with mitigating \emph{threats to validity}?
To answer this question, we consider each of the usual kinds of threats (construct, conclusion, internal, and external), and assess them for the case studies in \autoref{sec:case-studies}.

Bayesian analysis is unlikely to affect \emph{construct} validity, which has to do with whether we measured what the study was supposed to measure. 
This threat is very limited for the programming language and testing studies (\autoref{sec:rosetta} and \autoref{sec:testing}), which target well-defined and understood measures (running time, number of bugs).
It is potentially more significant for the agile vs.\ structured study, because classifying processes in only two categories (agile and structured) may be partly fuzzy and subjective; however, \autoref{sec:agile} discusses how the analysis is quite robust w.r.t.\ how this classification is done, which gives us confidence in its results.

\emph{Conclusion} validity depends on the application of appropriate statistical tests.
As we discuss in \autoref{sec:classic-vs-bayes}, frequentist hypothesis testing techniques are questionable because they do not properly assess significance; switching to Bayesian analysis can certainly help in this respect.
Thus, conclusion validity threats are lower in our three case studies than in the original studies that provided the data.

\emph{Internal validity} is mainly concerned with whether causality is correctly evaluated.
This depends on several details of experimental design that are generally independent of whether frequentist or Bayesian statistics are used.
One important aspect of internal validity pertains to the avoidance of \emph{bias}; this is where Bayesian statistics can help, thanks to its ability of weighting out many different competing models rather than restricting the analysis to two predefined hypotheses (null vs.\ alternative hypothesis).
This aspect is particularly relevant for the agile vs.\ structured study in the way it uses Bayes factors. 

Since it integrates previous, or otherwise independently obtained, information in the form or priors, Bayesian analysis can help mitigate threats to \emph{external validity}, which concern the generalizability of findings.
Using an informative prior makes the statistics reflect not just the current experimental data but also prior knowledge and assumptions on the subject; conversely, being able to get to the same conclusions using different, uninformative priors indicates that the experimental evidence is strong over initial assumptions.
In both cases, Bayesian statistics support analyses where generalizability is more explicitly taken into account instead of being just an afterthought.
This applies to all three case studies, and in particular to the programming language performance analysis (\autoref{sec:rosetta}) which integrated independent information to boost the confidence in the results.

\section{Practical Guidelines}
\label{sec:conclusions}

We conclude by summarizing practical guidelines to perform Bayesian analysis on empirical data from diverse software-engineering research.

\begin{itemize}
\item
If previous studies on the same subject are available, consider incorporating their data into the analysis in the form of prior---if only to estimate to what extent the interpretation of the new results changes according to what prior is used.

\item
To allow other researchers to do the same with your data, make it available in machine-readable form in addition to statistics and visualizations.

\item
Try to compute distributions of estimates rather than only single-point estimates. Visualize data as well as the computed distributions, and use the visual information to direct and refine your analysis.

\item
Consider alternatives to statistical hypothesis testing, for example the computation of Bayes factors; in any case, do not rely solely on the $p$-value to draw conclusions.

\item
More generally, avoid phrasing your analysis in terms of binary antithetical choices.
No statistical tests can substitute careful, informed modeling of assumptions. 
\end{itemize}

\clearpage


\iflong
\clearpage
\newpage
\onecolumn
\appendix
\newcommand{\TestingDataDir}{.}
\newcommand{\AgileDataDir}{.}
\newcommand{\RosettaDataDir}{.}
\def\GraphWidth{0.4\textwidth}
\def\AllGraphWidth{0.5\textwidth}

\newcommand{\NetworkFig}[3][]{%
\begin{figure}[!h]
\centering
  \begin{tikzpicture}[
  lang/.style={draw=none,font=\footnotesize,inner sep=1pt,outer sep=1pt},
  align=center, xscale=0.4, yscale=0.5
  ]
  \input{\RosettaDataDir/network_#2.tex}
  \end{tikzpicture}
\caption{#3}
\label{fig:#2_network}
\end{figure}
}

\newcommand{\GraphC}[2]{%
\begin{figure}[!p]
\begin{center}
\begin{tabular}{cc}
\includegraphicsifexists[width=\GraphWidth]{\RosettaDataDir/#1_C-Csharp.pdf}
&
\includegraphicsifexists[width=\GraphWidth]{\RosettaDataDir/#1_C-Fsharp.pdf} \\
\includegraphicsifexists[width=\GraphWidth]{\RosettaDataDir/#1_C-Go.pdf}
&
\includegraphicsifexists[width=\GraphWidth]{\RosettaDataDir/#1_C-Haskell.pdf} \\
\includegraphicsifexists[width=\GraphWidth]{\RosettaDataDir/#1_C-Java.pdf}
&
\includegraphicsifexists[width=\GraphWidth]{\RosettaDataDir/#1_C-Python.pdf} \\
\includegraphicsifexists[width=\GraphWidth]{\RosettaDataDir/#1_C-Ruby.pdf}
\end{tabular}
\end{center}
\caption{#2}
\label{fig:#1:C}
\end{figure}%
}

\newcommand{\GraphCsharp}[2]{%
\begin{figure}[!p]
\begin{center}
\begin{tabular}{cc}
\includegraphicsifexists[width=\GraphWidth]{\RosettaDataDir/#1_Csharp-Fsharp.pdf}
&
\includegraphicsifexists[width=\GraphWidth]{\RosettaDataDir/#1_Csharp-Go.pdf} \\
\includegraphicsifexists[width=\GraphWidth]{\RosettaDataDir/#1_Csharp-Haskell.pdf}
&
\includegraphicsifexists[width=\GraphWidth]{\RosettaDataDir/#1_Csharp-Java.pdf} \\
\includegraphicsifexists[width=\GraphWidth]{\RosettaDataDir/#1_Csharp-Python.pdf}
&
\includegraphicsifexists[width=\GraphWidth]{\RosettaDataDir/#1_Csharp-Ruby.pdf}
\end{tabular}
\end{center}
\caption{#2}
\label{fig:#1:Csharp}
\end{figure}%
}

\newcommand{\GraphFSharp}[2]{%
\begin{figure}[!p]
\begin{center}
\begin{tabular}{cc}
\includegraphicsifexists[width=\GraphWidth]{\RosettaDataDir/#1_Fsharp-Go.pdf}
&
\includegraphicsifexists[width=\GraphWidth]{\RosettaDataDir/#1_Fsharp-Haskell.pdf} \\
\includegraphicsifexists[width=\GraphWidth]{\RosettaDataDir/#1_Fsharp-Java.pdf}
&
\includegraphicsifexists[width=\GraphWidth]{\RosettaDataDir/#1_Fsharp-Python.pdf} \\
\includegraphicsifexists[width=\GraphWidth]{\RosettaDataDir/#1_Fsharp-Ruby.pdf}
\end{tabular}
\end{center}
\caption{#2}
\label{fig:#1:Fsharp}
\end{figure}%
}

\newcommand{\GraphGo}[2]{%
\begin{figure}[!p]
\begin{center}
\begin{tabular}{cc}
\includegraphicsifexists[width=\GraphWidth]{\RosettaDataDir/#1_Go-Haskell.pdf}
&
\includegraphicsifexists[width=\GraphWidth]{\RosettaDataDir/#1_Go-Java.pdf} \\
\includegraphicsifexists[width=\GraphWidth]{\RosettaDataDir/#1_Go-Python.pdf}
&
\includegraphicsifexists[width=\GraphWidth]{\RosettaDataDir/#1_Go-Ruby.pdf}
\end{tabular}
\end{center}
\caption{#2}
\label{fig:#1:Go}
\end{figure}%
}

\newcommand{\GraphHaskell}[2]{%
\begin{figure}[!p]
\begin{center}
\begin{tabular}{cc}
\includegraphicsifexists[width=\GraphWidth]{\RosettaDataDir/#1_Haskell-Java.pdf}
&
\includegraphicsifexists[width=\GraphWidth]{\RosettaDataDir/#1_Haskell-Python.pdf} \\
\includegraphicsifexists[width=\GraphWidth]{\RosettaDataDir/#1_Haskell-Ruby.pdf}
\end{tabular}
\end{center}
\caption{#2}
\label{fig:#1:Haskell}
\end{figure}%
}

\newcommand{\GraphJava}[2]{%
\begin{figure}[!p]
\begin{center}
\begin{tabular}{cc}
\includegraphicsifexists[width=\GraphWidth]{\RosettaDataDir/#1_Java-Python.pdf}
&
\includegraphicsifexists[width=\GraphWidth]{\RosettaDataDir/#1_Java-Ruby.pdf}
\end{tabular}
\end{center}
\caption{#2}
\label{fig:#1:Java}
\end{figure}%
}

\newcommand{\GraphPython}[2]{%
\begin{figure}[!p]
\begin{center}
\begin{tabular}{cc}
\includegraphicsifexists[width=\GraphWidth]{\RosettaDataDir/#1_Python-Ruby.pdf}
\end{tabular}
\end{center}
\caption{#2}
\label{fig:#1:Python}
\end{figure}%
}

\newcommand{\GraphRosetta}[2]{%
\GraphC{#1}{#2 (C vs.\ other languages)}
\GraphCsharp{#1}{#2 (C\# vs.\ other languages)}
\GraphFSharp{#1}{#2 (F\# vs.\ other languages)}
\GraphGo{#1}{#2 (Go vs.\ other languages)}
\GraphHaskell{#1}{#2 (Haskell vs.\ other languages)}
\GraphJava{#1}{#2 (Java vs.\ other languages)}
\GraphPython{#1}{#2 (Python  vs.\ other languages)}
}

\newcommand{\GraphNBugsPrior}[3]{%
\clearpage
\begin{figure}[!p]
\begin{center}
\begin{tabular}{cc}
\includegraphicsifexists[width=\GraphWidth]{\TestingDataDir/#1_M_1_#2.pdf}
&
\includegraphicsifexists[width=\GraphWidth]{\TestingDataDir/#1_M_2_#2.pdf} \\
\includegraphicsifexists[width=\GraphWidth]{\TestingDataDir/#1_M_3_#2.pdf}
&
\includegraphicsifexists[width=\GraphWidth]{\TestingDataDir/#1_M_4_#2.pdf} \\
\includegraphicsifexists[width=\GraphWidth]{\TestingDataDir/#1_M_5_#2.pdf}
&
\includegraphicsifexists[width=\GraphWidth]{\TestingDataDir/#1_M_6_#2.pdf} \\
\includegraphicsifexists[width=\GraphWidth]{\TestingDataDir/#1_M_7_#2.pdf}
\end{tabular}
\end{center}
\caption{#3}
\label{fig:#1:#2-M1-7}
\end{figure}%

\clearpage
\begin{figure}[!p]
\begin{center}
\begin{tabular}{cc}
\includegraphicsifexists[width=\GraphWidth]{\TestingDataDir/#1_M_8_#2.pdf}
&
\includegraphicsifexists[width=\GraphWidth]{\TestingDataDir/#1_M_9_#2.pdf} \\
\includegraphicsifexists[width=\GraphWidth]{\TestingDataDir/#1_M_10_#2.pdf}
&
\includegraphicsifexists[width=\GraphWidth]{\TestingDataDir/#1_M_11_#2.pdf} \\
\includegraphicsifexists[width=\GraphWidth]{\TestingDataDir/#1_M_12_#2.pdf}
&
\includegraphicsifexists[width=\GraphWidth]{\TestingDataDir/#1_M_13_#2.pdf} \\
\includegraphicsifexists[width=\GraphWidth]{\TestingDataDir/#1_M_14_#2.pdf}
\end{tabular}
\end{center}
\caption{#3}
\label{fig:#1:#2-M8-14}
\end{figure}%

\clearpage
\begin{figure}[!p]
\begin{center}
\begin{tabular}{cc}
\includegraphicsifexists[width=\GraphWidth]{\TestingDataDir/#1_M_15_#2.pdf}
&
\includegraphicsifexists[width=\GraphWidth]{\TestingDataDir/#1_M_16_#2.pdf} \\
\includegraphicsifexists[width=\GraphWidth]{\TestingDataDir/#1_M_17_#2.pdf}
&
\includegraphicsifexists[width=\GraphWidth]{\TestingDataDir/#1_M_18_#2.pdf} \\
\includegraphicsifexists[width=\GraphWidth]{\TestingDataDir/#1_M_19_#2.pdf}
&
\includegraphicsifexists[width=\GraphWidth]{\TestingDataDir/#1_M_20_#2.pdf} \\
\includegraphicsifexists[width=\GraphWidth]{\TestingDataDir/#1_M_21_#2.pdf}
\end{tabular}
\end{center}
\caption{#3}
\label{fig:#1:#2-M15-21}
\end{figure}%
}

\newcommand{\GraphNBugs}[2]{%
\GraphNBugs{#1}{uniform}{#2}
\GraphNBugs{#1}{jeffreys}{#2}
}

\newcommand{\PlotsForPriors}[2]{%
\begin{figure}[!h]
\begin{center}
\begin{subfigure}[t]{\AllGraphWidth}
\centering
\includegraphicsifexists[width=\textwidth]{\TestingDataDir/#1_uniform.pdf}
\end{subfigure}%
~%
\begin{subfigure}[t]{\AllGraphWidth}
\centering
\includegraphicsifexists[width=\textwidth]{\TestingDataDir/#1_jeffreys.pdf}
\end{subfigure}
\end{center}
\caption{#2}
\label{fig:#1-all-priors}
\end{figure}%
}

\tableofcontents
\listoffigures
\listoftables
\pagebreak

\subsection{Rosetta Code Study} \label{app:rosetta}

The tables comparing programming languages gray out cells corresponding to values that may provide lower confidence.
Given a confidence intervals $(c, C)$, mean $\mu$, and median $m$, let $\delta = C - c \geq 0$ be the interval size and $s = \min(|c|, |C|)$ be the absolute value of the endpoint closer to the origin.
If $c < 0 < C$ or $-1.1 < \mu < 1.1$ the comparison may be not significant (dark gray background); if the comparison is significant and $\delta \geq s$, $s \leq 2$, and $|m| \leq 2$ the comparison may be only weakly significant (light gray background); in all other cases it is significant (no gray background).

\subsubsection{Running time}

\begin{table}[ht]
\centering
\begingroup\scriptsize
\begin{tabular}{cc|rrrrrrr}
  \hline
\textsc{language} & \textsc{measure} & \multicolumn{1}{c}{C} & \multicolumn{1}{c}{C\#} & \multicolumn{1}{c}{F\#} & \multicolumn{1}{c}{Go} & \multicolumn{1}{c}{Haskell} & \multicolumn{1}{c}{Java} & \multicolumn{1}{c}{Python} \\ 
  \hline
C\# & CI & \strongsig{(-10.1, -8.7)} &  &  &  &  &  &  \\ 
   & $m$ & -9.22 &  &  &  &  &  &  \\ 
   & $\mu$ & -9.57 &  &  &  &  &  &  \\ 
   \hline
F\# & CI & \strongsig{(-76.0, -64.5)} & \strongsig{(-8.9, -4.6)} &  &  &  &  &  \\ 
   & $m$ & -72.61 & -5.29 &  &  &  &  &  \\ 
   & $\mu$ & -70.58 & -6.83 &  &  &  &  &  \\ 
   \hline
Go & CI & \strongsig{(-2.3, -1.3)} & \weaksig{(1.0, 2.5)} & \strongsig{(16.9, 20.6)} &  &  &  &  \\ 
   & $m$ & -1.67 & 1.15 & 18.21 &  &  &  &  \\ 
   & $\mu$ & -1.77 & 1.52 & 18.78 &  &  &  &  \\ 
   \hline
Haskell & CI & \strongsig{(-5.9, -5.7)} & \strongsig{(1.2, 1.7)} & \strongsig{(3.2, 15.5)} & \strongsig{(2.4, 2.5)} &  &  &  \\ 
   & $m$ & -5.76 & 1.23 & 6.77 & 2.49 &  &  &  \\ 
   & $\mu$ & -5.82 & 1.35 & 7.03 & 2.47 &  &  &  \\ 
   \hline
Java & CI & \strongsig{(-3.2, -3.1)} & \strongsig{(-2.0, -1.3)} & \strongsig{(5.7, 7.5)} & \strongsig{(-8.5, -7.6)} & \strongsig{(-8.6, -8.3)} &  &  \\ 
   & $m$ & -3.18 & -1.77 & 6.94 & -8.02 & -8.63 &  &  \\ 
   & $\mu$ & -3.18 & -1.68 & 6.71 & -8.06 & -8.39 &  &  \\ 
   \hline
Python & CI & \strongsig{(-54.2, -32.3)} & \notsig{(-1.4, 1.8)} & \strongsig{(2.1, 12.7)} & \strongsig{(-27.2, -17.7)} & \weaksig{(-5.0, -1.2)} & \notsig{(-2.1, 2.2)} &  \\ 
   & $m$ & -52.47 & 1.3 & 7.93 & -23.01 & -1.82 & 1.76 &  \\ 
   & $\mu$ & -44.23 & 1.31 & 7 & -22.78 & -2.58 & 1.31 &  \\ 
   \hline
Ruby & CI & \strongsig{(-124.0, -90.9)} & \strongsig{(-21.1, -11.6)} & \notsig{(1.0, 1.1)} & \strongsig{(-142.2, -36.9)} & \strongsig{(-22.0, -19.9)} & \strongsig{(-17.0, -8.6)} & \strongsig{(-19.7, -14.0)} \\ 
   & $m$ & -100.69 & -16.96 & 1.05 & -141.92 & -21.85 & -15.32 & -15.03 \\ 
   & $\mu$ & -105.99 & -16.42 & 1.05 & -105.32 & -20.51 & -12.76 & -16.8 \\ 
   \hline
\end{tabular}
\endgroup
\caption{Comparison of running times} 
\label{tab:tab_runtime}
\end{table}

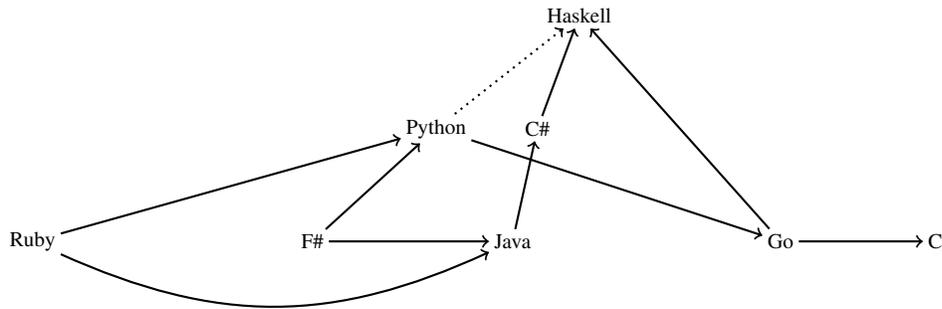
\begin{figure}[!h]
\centering
  \begin{tikzpicture}[
  lang/.style={draw=none,font=\footnotesize,inner sep=1pt,outer sep=1pt},
  align=center, xscale=0.4, yscale=0.5
  ]
  \node [lang] (C) at (15,0) {C};
\node [lang] (C sharp) at (1.78079763207794,3) {C\#};
\node [lang] (F sharp) at (-5.70081859805448,0) {F\#};
\node [lang] (Go) at (9.85948170872709,0) {Go};
\node [lang] (Haskell) at (3.17010190081243,6) {Haskell};
\node [lang] (Java) at (0.952487397289838,0) {Java};
\node [lang] (Python) at (-1.58840396503615,3) {Python};
\node [lang] (Ruby) at (-15,0) {Ruby};
\draw [->,thick,colorarrow](Go) edge (C);
\draw [->,thick,colorarrow](C sharp) edge (Haskell);
\draw [->,thick,colorarrow](Java) edge (C sharp);
\draw [->,thick,colorarrow](F sharp) edge (Java);
\draw [->,thick,colorarrow](F sharp) edge (Python);
\draw [->,thick,colorarrow](Go) edge (Haskell);
\draw [->,thick,colorarrow](Python) edge (Go);
\draw [->,dotted,thick,colorarrow](Python) edge (Haskell);
\draw [->,thick,colorarrow,bend right=20](Ruby) edge (Java);
\draw [->,thick,colorarrow](Ruby) edge (Python);
  \end{tikzpicture}
\caption{Running time}
\label{fig:runtime_network}
\end{figure}

\GraphRosetta{runtime}{Running time}

\clearpage
\subsubsection{Memory usage}

\rosetta's \emph{memory} experiments measure the maximum RAM usage of programs implementing the same tasks: $M(\ell, t)$ is the maximum RAM used by the \emph{best} (that is, using the least amount of memory) implementation of $t$ in language $\ell$; and $\vec{M}(\ell_1, \ell_2)$ is the vector of elementwise ratios \eqref{eq:ratio-def}, defined like $\vec{S}(\ell_1, \ell_2)$ but with respect to $\vec{M}(\ell_1)$ and $\vec{M}(\ell_2)$.
\bench also includes \emph{memory} experiments that determine vectors $\bvec{M}(\ell_1, \ell_2)$ and sets $\bvec{M}_{\Delta}(\ell_1, \ell_2, t)$, defined just like their counterparts for running time but measuring the maximum RAM used in each case.
The following graphs, tables, and plots are the counterpart of the analysis of running time $S$ with respect to the data $M$ on memory usage.

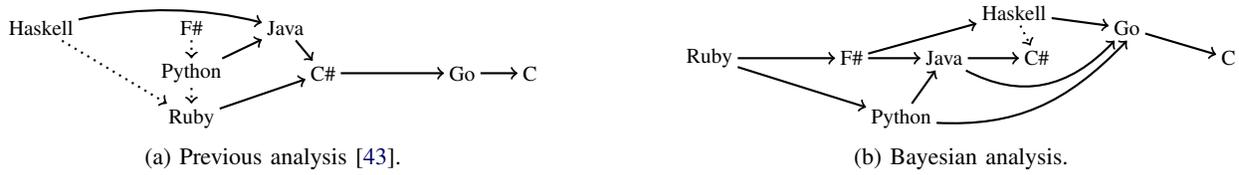
\begin{figure*}[!t]
\centering
\begin{subfigure}[t]{0.5\textwidth}
\centering
  \begin{tikzpicture}[
  lang/.style={draw=none,font=\footnotesize,inner sep=1pt,outer sep=1pt},
  align=center, xscale=0.5, yscale=0.2
  ]
  \node [lang] (C) at (0,0) {C};
  \node [lang] (Go) at (-1.8,0) {Go};
  \node [lang] (C-sharp) at (-5.5,0) {C\#};
  \node [lang] (F-Sharp) at (-9,3) {F\#};
  \node [lang] (Python) at (-9,0) {Python};
  \node [lang] (Ruby) at (-9,-3) {Ruby};
  \node [lang] (Java) at (-6.5,3) {Java};
  \node [lang] (Haskell) at (-13,3) {Haskell};
 
  \draw [->,thick](Go) edge (C);
  \draw [->,dotted,thick](F-Sharp) edge (Python);
  \draw [->,thick](C-sharp) edge (Go);
  \draw [->,thick](Java) edge (C-sharp);
  \draw [->,thick](Ruby) edge (C-sharp);
  \draw [->,thick,bend left](Haskell) edge (Java);
  \draw [->,dotted,thick](Haskell) edge (Ruby);
  \draw [->,thick](Python) edge (Java);
  \draw [->,dotted,thick](Python) edge (Ruby);
\end{tikzpicture}
\caption{Previous analysis~\cite{rosetta}.}
\label{fig:OLD-network_memory}
\end{subfigure}%
~%
\begin{subfigure}[t]{0.5\textwidth}
\centering
  \begin{tikzpicture}[
  lang/.style={draw=none,font=\footnotesize,inner sep=1pt,outer sep=1pt},
  align=center, xscale=0.3, yscale=0.4
  ]
\node [lang] (C) at (13,0) {C};
\node [lang] (C sharp) at (4.5,0) {C\#};
\node [lang] (F sharp) at (-3.7,0) {F\#};
\node [lang] (Go) at (8.5,1) {Go};
\node [lang] (Haskell) at (3.5,1.5) {Haskell};
\node [lang] (Java) at (0.4,0) {Java};
\node [lang] (Python) at (-1.5,-2) {Python};
\node [lang] (Ruby) at (-10,0) {Ruby};
\draw [->,thick,colorarrow](Go) edge (C);
\draw [->,thick,colorarrow](Haskell) edge (Go);
\draw [->,thick,colorarrow,bend right](Java) edge (Go);
\draw [->,thick,colorarrow](F sharp) edge (Haskell);
\draw [->,thick,colorarrow,bend right=20](Python) edge (Go.south);
\draw [->,thick,dotted,colorarrow](Haskell) edge (C sharp);
\draw [->,thick,colorarrow](Java) edge (C sharp);
\draw [->,thick,colorarrow](F sharp) edge (Java);
\draw [->,thick,colorarrow](Ruby) edge (F sharp);
\draw [->,thick,colorarrow](Ruby) edge (Python);
\draw [->,thick,colorarrow](Python) edge (Java);
  \end{tikzpicture}
\caption{Bayesian analysis.}
\label{fig:network_memory}
\end{subfigure}
\label{fig:both_network_memory}
\caption{Comparison of used memory: qualitative summaries.}
\end{figure*}

\begin{table}[ht]
\centering
\begingroup\scriptsize
\begin{tabular}{cc|rrrrrrr}
  \hline
\textsc{language} & \textsc{measure} & \multicolumn{1}{c}{C} & \multicolumn{1}{c}{C\#} & \multicolumn{1}{c}{F\#} & \multicolumn{1}{c}{Go} & \multicolumn{1}{c}{Haskell} & \multicolumn{1}{c}{Java} & \multicolumn{1}{c}{Python} \\ 
  \hline
C\# & CI & \notsig{(-12.9, 9.6)} &  &  &  &  &  &  \\ 
   & $m$ & 9.83 &  &  &  &  &  &  \\ 
   & $\mu$ & 1.67 &  &  &  &  &  &  \\ 
   \hline
F\# & CI & \strongsig{(-65.3, -22.1)} & \strongsig{(-9.2, -1.4)} &  &  &  &  &  \\ 
   & $m$ & -35.28 & -6.32 &  &  &  &  &  \\ 
   & $\mu$ & -39.51 & -5.19 &  &  &  &  &  \\ 
   \hline
Go & CI & \strongsig{(-3.2, -1.9)} & \notsig{(-7.0, 8.8)} & \strongsig{(61.6, 63.2)} &  &  &  &  \\ 
   & $m$ & -1.84 & -2.96 & 62.74 &  &  &  &  \\ 
   & $\mu$ & -2.39 & 0.68 & 62.72 &  &  &  &  \\ 
   \hline
Haskell & CI & \strongsig{(-12.8, -8.5)} & \strongsig{(-5.0, -1.0)} & \strongsig{(46.1, 49.3)} & \strongsig{(-3.6, -3.4)} &  &  &  \\ 
   & $m$ & -11.11 & -4.14 & 48.76 & -3.56 &  &  &  \\ 
   & $\mu$ & -10.71 & -2.74 & 48.05 & -3.56 &  &  &  \\ 
   \hline
Java & CI & \notsig{(-7.8, 1.7)} & \strongsig{(-4.2, -1.2)} & \strongsig{(1.4, 4.1)} & \strongsig{(-33.5, -31.8)} & \notsig{(-2.7, 1.2)} &  &  \\ 
   & $m$ & -1.06 & -2.34 & 2.26 & -32.9 & -2.07 &  &  \\ 
   & $\mu$ & -3.4 & -2.52 & 2.3 & -32.89 & -1.44 &  &  \\ 
   \hline
Python & CI & \strongsig{(-18.0, -2.1)} & \notsig{(-15.5, 24.8)} & \notsig{(-4.3, 7.1)} & \strongsig{(-7.0, -1.5)} & \notsig{(-2.2, 4.6)} & \strongsig{(-10.6, -1.6)} &  \\ 
   & $m$ & -11.1 & -7.58 & -2.62 & -6.82 & 2.35 & -3.51 &  \\ 
   & $\mu$ & -9.73 & 0.68 & 1.83 & -4.49 & 2.23 & -5.73 &  \\ 
   \hline
Ruby & CI & \strongsig{(-229.5, -157.0)} & \strongsig{(-36.5, -6.2)} & \strongsig{(-23.6, -2.9)} & \strongsig{(-194.8, -174.3)} & \strongsig{(-60.7, -10.6)} & \strongsig{(-25.6, -5.3)} & \strongsig{(-12.2, -4.3)} \\ 
   & $m$ & -154.78 & -15.95 & -18.33 & -190.44 & -43.66 & -13 & -10.82 \\ 
   & $\mu$ & -195.45 & -17.16 & -9.92 & -184.58 & -35.02 & -12.07 & -7.18 \\ 
   \hline
\end{tabular}
\endgroup
\caption{Comparison of maximum RAM usage} 
\label{tab:tab_memory}
\end{table}

\begin{figure}[!h]
\centering
  \begin{tikzpicture}[
  lang/.style={draw=none,font=\footnotesize,inner sep=1pt,outer sep=1pt},
  align=center, xscale=0.4, yscale=0.5
  ]
  \input{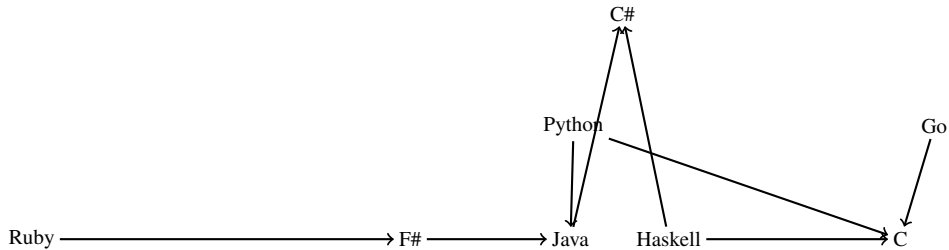}
  \end{tikzpicture}
\caption{Maximum RAM usage}
\label{fig:memory_network}
\end{figure}

\GraphRosetta{memory}{Maximum RAM usage}

\clearpage
\pagebreak
\subsection{Agile vs.\ Structured Study} \label{app:agile}

\subsubsection{Success}

\autoref{tab:agile-outcome-distros} shows the distributions $\outcome_C$ for the nine combinations of process categories we analyze: $A$, $AIL$, $AILT$, $AIT$, $AL$, $ALT$, $AT$, $AT$, $IT$, and $T$.
These are used as baseline distributions in computing the Bayes factors.
\autoref{tab:agile-outcome-distros} is like \autoref{tab:agile-distros} in the main paper.

\autoref{tab:agile-bayes-factors-outcome} shows the Bayes factors $K_C(\overline{D})$ from \agile, for each choice of baseline distribution $C$ in \autoref{tab:agile-outcome-distros}.
Each row uses different weights $w_p$s for the likelihoods \eqref{eq:like-agile-ha} and \eqref{eq:like-agile-h0}.
The table is partitioned in four parts separated by horizontal lines; they correspond, from top to bottom, to the analysis using different choices of lower bound $b = 1, 3, 4, 5$ for the scaling of data (row header). 
The scaling of data translates a project outcome $1 \leq x \leq 10$  into an outcome $0 \leq x' \leq r$ using lower bound $b$ as follows.
For $0 \leq k \leq r$, let $\sigma_r(k) = b + k (10 - b)/r$ be a uniformly spaced point over $[b,10]$; then $x' = \argmin_k |\sigma_r(k) - x|$.
In the main paper, \autoref{tab:agile-bayes-factors-summary} shows the data for $b = 1$.

\begin{table}[ht]
\centering
\begin{tabular}{c|*{9}{r}}
& \multicolumn{1}{c}{$A$}
& \multicolumn{1}{|c}{$AIL$}
& \multicolumn{1}{|c}{$AILT$}
& \multicolumn{1}{|c}{$AIT$}
& \multicolumn{1}{|c}{$AL$} 
& \multicolumn{1}{|c}{$ALT$}
& \multicolumn{1}{|c}{$AT$} 
& \multicolumn{1}{|c}{$IT$}
& \multicolumn{1}{|c}{$T$}
\\
\hline
$\outcome_{C}[{0}]$  &  7{\scriptsize\,\%}  &  8{\scriptsize\,\%}  &  10{\scriptsize\,\%}  &  11{\scriptsize\,\%}  &  7{\scriptsize\,\%}  &  11{\scriptsize\,\%}  &  12{\scriptsize\,\%}  &  12{\scriptsize\,\%}  &  18{\scriptsize\,\%} \\
$\outcome_{C}[{1}]$  &  30{\scriptsize\,\%}  &  27{\scriptsize\,\%}  &  28{\scriptsize\,\%}  &  29{\scriptsize\,\%}  &  27{\scriptsize\,\%}  &  29{\scriptsize\,\%}  &  31{\scriptsize\,\%}  &  29{\scriptsize\,\%}  &  32{\scriptsize\,\%} \\
$\outcome_{C}[{2}]$  &  63{\scriptsize\,\%}  &  65{\scriptsize\,\%}  &  62{\scriptsize\,\%}  &  60{\scriptsize\,\%}  &  66{\scriptsize\,\%}  &  60{\scriptsize\,\%}  &  57{\scriptsize\,\%}  &  59{\scriptsize\,\%}  &  50{\scriptsize\,\%} \\
\end{tabular}
\caption{Different distributions for the baseline probability of \emph{outcome} for software projects: for $C$ one of the subsets of $\{A, I, L, T\}$ in the column headers, $\outcome_C[k]$ is the probability that outcome is $k \in \{0, 1, 2\}$ in processes of category $C$.}
\label{tab:agile-outcome-distros}
\end{table}

\begin{table}[ht]
\centering
\begin{tabular}{l|*{9}{r}}
& \multicolumn{1}{c}{$A$}
& \multicolumn{1}{|c}{$AIL$}
& \multicolumn{1}{|c}{$AILT$}
& \multicolumn{1}{|c}{$AIT$}
& \multicolumn{1}{|c}{$AL$} 
& \multicolumn{1}{|c}{$ALT$}
& \multicolumn{1}{|c}{$AT$} 
& \multicolumn{1}{|c}{$IT$}
& \multicolumn{1}{|c}{$T$}
\\
\hline
uniform  &  0.2456  &  0.2609  &  0.1655  &  0.1362  &  0.2881  &  0.1228  &  0.0780  &  0.0986  &  0.0131 \\
triangle  &  0.2472  &  0.2614  &  0.1704  &  0.1411  &  0.2883  &  0.1284  &  0.0827  &  0.1037  &  0.0151 \\
power  &  0.2484  &  0.2618  &  0.1735  &  0.1441  &  0.2885  &  0.1320  &  0.0855  &  0.1069  &  0.0162 \\
exp  &  0.2525  &  0.2633  &  0.1860  &  0.1567  &  0.2891  &  0.1471  &  0.0986  &  0.1208  &  0.0229 \\
\hline
uniform  &  0.0329  &  0.0277  &  0.0625  &  0.0735  &  0.0188  &  0.0782  &  0.0888  &  0.0853  &  0.0424 \\
triangle  &  0.0348  &  0.0295  &  0.0648  &  0.0760  &  0.0201  &  0.0807  &  0.0915  &  0.0879  &  0.0460 \\
power  &  0.0361  &  0.0307  &  0.0665  &  0.0777  &  0.0210  &  0.0824  &  0.0933  &  0.0897  &  0.0480 \\
exp  &  0.0414  &  0.0358  &  0.0726  &  0.0843  &  0.0248  &  0.0888  &  0.1003  &  0.0965  &  0.0583 \\
\hline
uniform  &  0.0406  &  0.0336  &  0.0854  &  0.1055  &  0.0220  &  0.1151  &  0.1461  &  0.1325  &  0.1102 \\
triangle  &  0.0427  &  0.0356  &  0.0877  &  0.1077  &  0.0235  &  0.1170  &  0.1478  &  0.1344  &  0.1156 \\
power  &  0.0442  &  0.0370  &  0.0894  &  0.1094  &  0.0245  &  0.1185  &  0.1491  &  0.1358  &  0.1187 \\
exp  &  0.0501  &  0.0427  &  0.0953  &  0.1152  &  0.0287  &  0.1235  &  0.1537  &  0.1406  &  0.1330 \\
\hline
uniform  &  0.0406  &  0.0336  &  0.0854  &  0.1055  &  0.0220  &  0.1151  &  0.1461  &  0.1325  &  0.1102 \\
triangle  &  0.0427  &  0.0356  &  0.0877  &  0.1077  &  0.0235  &  0.1170  &  0.1478  &  0.1344  &  0.1156 \\
power  &  0.0442  &  0.0370  &  0.0894  &  0.1094  &  0.0245  &  0.1185  &  0.1491  &  0.1358  &  0.1187 \\
exp  &  0.0501  &  0.0427  &  0.0953  &  0.1152  &  0.0287  &  0.1235  &  0.1537  &  0.1406  &  0.1330 \\
\end{tabular}
\caption{Bayes factors $K$ estimating to what extent the hypothesis ``agile leads to more successful projects'' is supported over the other hypothesis ``agile is no better than structured''.} 
\label{tab:agile-bayes-factors-outcome}
\end{table}

\subsubsection{Importance for customers}

The data in \agile also reports, for each project $p \in P$, its \emph{importance} for customers $I(p)$ on a scale 1--10, where 1 denotes an unimportant project and 10 denotes a very critical one.
The multisets $I_A$, $I_S$, and $I$ are defined similarly to $O_A$, $O_S$, and $O$ but for importance assessments.

The data in \itproj also assesses ``stakeholder value'', which we can assimilate to \emph{importance} for customers---a most significant group of stakeholders.
For each category $c \in \{A, H, I, L, T\}$, each respondent $r \in R$ assesses the projects in category $c$ according to their stakeholder value $v_c(r)$ on a scale 0--4, where 0 denotes no value and 4 denotes very high value.\footnote{Respondents could also mark this question as ``not applicable'', in which case we ignored their answer.}
For every non-empty subset $C \subseteq \{A, H, I, L, T\}$, we define a distribution $\stakeholders_C$ over values in the range $[0..4]$ as follows.
If $C$ is a singleton set $\{c\}$ with $c \in \{A, H, I, L, T\}$, $\stakeholders_c[k]$ is the probability $|\{r \in R \mid v_c(r) = k\}|/|R|$ that a project in category $c$ has value $k$.
If $C$ is any non-empty subset of $\{A, H, I, L, T\}$, $\stakeholders_C[k]$ is the weighted average $\stakeholders_C[k] = \sum_{c \in C}\stakeholders_{c}[k]/|C|$.
\autoref{tab:agile-stakeholders-distros} shows the distributions $\stakeholders_C$ for the nine combinations of process categories we analyze.

\begin{table}[ht]
\centering
\begin{tabular}{c|*{9}{r}}
\hline
& \multicolumn{1}{c}{$A$}
& \multicolumn{1}{c}{$AIL$}
& \multicolumn{1}{c}{$AILT$}
& \multicolumn{1}{c}{$AIT$}
& \multicolumn{1}{c}{$AL$} 
& \multicolumn{1}{c}{$ALT$}
& \multicolumn{1}{c}{$AT$} 
& \multicolumn{1}{c}{$IT$}
& \multicolumn{1}{c}{$T$}
\\
\hline
$\stakeholders_{C}[{0}]$  &  0{\scriptsize\,\%}  &  1{\scriptsize\,\%}  &  2{\scriptsize\,\%}  &  2{\scriptsize\,\%}  &  0{\scriptsize\,\%}  &  3{\scriptsize\,\%}  &  3{\scriptsize\,\%}  &  3{\scriptsize\,\%}  &  7{\scriptsize\,\%} \\
$\stakeholders_{C}[{1}]$  &  4{\scriptsize\,\%}  &  3{\scriptsize\,\%}  &  9{\scriptsize\,\%}  &  9{\scriptsize\,\%}  &  6{\scriptsize\,\%}  &  14{\scriptsize\,\%}  &  15{\scriptsize\,\%}  &  11{\scriptsize\,\%}  &  26{\scriptsize\,\%} \\
$\stakeholders_{C}[{2}]$  &  15{\scriptsize\,\%}  &  18{\scriptsize\,\%}  &  20{\scriptsize\,\%}  &  21{\scriptsize\,\%}  &  13{\scriptsize\,\%}  &  19{\scriptsize\,\%}  &  21{\scriptsize\,\%}  &  24{\scriptsize\,\%}  &  28{\scriptsize\,\%} \\
$\stakeholders_{C}[{3}]$  &  30{\scriptsize\,\%}  &  35{\scriptsize\,\%}  &  35{\scriptsize\,\%}  &  35{\scriptsize\,\%}  &  33{\scriptsize\,\%}  &  34{\scriptsize\,\%}  &  32{\scriptsize\,\%}  &  37{\scriptsize\,\%}  &  35{\scriptsize\,\%} \\
$\stakeholders_{C}[{4}]$  &  51{\scriptsize\,\%}  &  43{\scriptsize\,\%}  &  34{\scriptsize\,\%}  &  33{\scriptsize\,\%}  &  48{\scriptsize\,\%}  &  30{\scriptsize\,\%}  &  29{\scriptsize\,\%}  &  25{\scriptsize\,\%}  &  4{\scriptsize\,\%} \\
\end{tabular}
\caption{Different distributions for the baseline probability of \emph{customer importance} for software projects: for $C$ one of the subsets of $\{A, I, L, T\}$ in the column headers, $\stakeholders_C[k]$ is the probability that customer importance is $k \in \{0, 1, 2\}$ in processes of category $C$.}
\label{tab:agile-stakeholders-distros}
\end{table}

The Bayes factors are computed just like the analysis of success mutatis mutandis.
\autoref{tab:agile-bayes-factors-summary} shows the Bayes factors for the data $I_A \cup I_S$ from \agile, using the same conventions as \autoref{tab:agile-bayes-factors-outcome} and using $\stakeholders_C$ as baseline distributions, for $C$ one of the nine distributions in \autoref{tab:agile-stakeholders-distros}.

\begin{table}[ht]
\centering
\begin{tabular}{l|*{9}{r}}
\hline
& \multicolumn{1}{c}{$A$}
& \multicolumn{1}{c}{$AIL$}
& \multicolumn{1}{c}{$AILT$}
& \multicolumn{1}{c}{$AIT$}
& \multicolumn{1}{c}{$AL$} 
& \multicolumn{1}{c}{$ALT$}
& \multicolumn{1}{c}{$AT$} 
& \multicolumn{1}{c}{$IT$}
& \multicolumn{1}{c}{$T$}
\\
\hline
uniform  &  0.0395  &  0.0494  &  0.0087  &  0.0087  &  0.0395  &  0.0038  &  0.0019  &  0.0019  &  0.0000 \\
triangle  &  0.0429  &  0.0549  &  0.0114  &  0.0116  &  0.0435  &  0.0057  &  0.0029  &  0.0030  &  0.0000 \\
power  &  0.0449  &  0.0570  &  0.0118  &  0.0121  &  0.0452  &  0.0058  &  0.0029  &  0.0030  &  0.0000 \\
exp  &  0.0547  &  0.0728  &  0.0217  &  0.0238  &  0.0566  &  0.0140  &  0.0074  &  0.0078  &  0.0001 \\
\hline
uniform  &  0.0003  &  0.0152  &  0.0800  &  0.0800  &  0.0003  &  0.0418  &  0.0205  &  0.0205  &  0.0001 \\
triangle  &  0.0003  &  0.0162  &  0.0839  &  0.0854  &  0.0003  &  0.0477  &  0.0238  &  0.0241  &  0.0002 \\
power  &  0.0004  &  0.0168  &  0.0855  &  0.0877  &  0.0004  &  0.0499  &  0.0248  &  0.0252  &  0.0001 \\
exp  &  0.0005  &  0.0196  &  0.0954  &  0.1024  &  0.0004  &  0.0679  &  0.0348  &  0.0365  &  0.0007 \\
\hline
uniform  &  0.0000  &  0.0000  &  0.0002  &  0.0002  &  0.0000  &  0.0053  &  0.0129  &  0.0129  &  0.0001 \\
triangle  &  0.0000  &  0.0000  &  0.0003  &  0.0003  &  0.0000  &  0.0063  &  0.0154  &  0.0155  &  0.0002 \\
power  &  0.0000  &  0.0000  &  0.0003  &  0.0003  &  0.0000  &  0.0067  &  0.0163  &  0.0164  &  0.0001 \\
exp  &  0.0000  &  0.0000  &  0.0004  &  0.0004  &  0.0000  &  0.0099  &  0.0242  &  0.0250  &  0.0007 \\
\hline
uniform  &  0.0000  &  0.0000  &  0.0010  &  0.0010  &  0.0000  &  0.0035  &  0.0203  &  0.0203  &  0.0007 \\
triangle  &  0.0000  &  0.0000  &  0.0012  &  0.0012  &  0.0000  &  0.0040  &  0.0231  &  0.0232  &  0.0011 \\
power  &  0.0000  &  0.0000  &  0.0014  &  0.0013  &  0.0000  &  0.0042  &  0.0247  &  0.0247  &  0.0011 \\
exp  &  0.0000  &  0.0000  &  0.0021  &  0.0020  &  0.0000  &  0.0057  &  0.0330  &  0.0331  &  0.0037 \\
\end{tabular}
\caption{Bayes factors $K$ estimating to what extent the hypothesis ``agile is used for projects that are more important for customers'' is supported over the other hypothesis ``agile is no better than structured''.}
\label{tab:agile-bayes-factors-stakeholders}
\end{table}

\clearpage
\subsection{Testing Study} \label{app:testing}

The uniform prior assigns constant probability to every pair $\alpha, beta$; Jeffreys prior~\cite{weibull-estimate} assigns it probability proportional to $(\alpha \beta)^{-1}$.

\begin{table}[!h]
\centering
\scriptsize
\setlength{\tabcolsep}{1pt}
\begin{adjustwidth}{-7mm}{-10mm}
\begin{tabular}{l| *{21}{r}}
& $C_1$  & $C_2$  & $C_3$  & $C_4$  & $C_5$  & $C_6$  & $C_7$  & $C_8$  & $C_9$  & $C_{10}$ & $C_{11}$ & $C_{12}$ & $C_{13}$ & $C_{14}$ & $C_{15}$ & $C_{16}$ & $C_{17}$ & $C_{18}$ & $C_{19}$ & $C_{20}$ & $C_{21}$ \\
\hline
$\mu(N)$  &  13  &  2  &  7  &  31  &  34  &  7  &  7  &  2  &  7  &  7  &  7  &  2  &  36  &  2  &  10  &  10  &  2  &  40  &  2  &  68  &  2 \\
$m(N)$  &  10  &  2  &  7  &  28  &  46  &  7  &  10  &  1  &  10  &  3  &  7  &  3  &  30  &  1  &  10  &  17  &  4  &  26  &  3  &  110  &  2 \\
$\widehat{N}$  &  9  &  0  &  4  &  26  &  26  &  4  &  4  &  0  &  4  &  4  &  4  &  0  &  26  &  0  &  6  &  6  &  0  &  40  &  0  &  60  &  0 \\
$N_{90}^l$  &  5  &  0  &  2  &  18  &  20  &  2  &  2  &  0  &  2  &  2  &  2  &  0  &  22  &  0  &  4  &  4  &  0  &  25  &  0  &  49  &  0 \\
$N_{90}^h$  &  26  &  7  &  17  &  54  &  57  &  17  &  17  &  7  &  17  &  17  &  17  &  7  &  60  &  7  &  22  &  22  &  7  &  66  &  7  &  100  &  7 \\
$m(N)/\textsc{loc}$  &  1.2E-02  &  1.2E-03  &  1.2E-02  &  1.4E-02  &  4.1E-02  &  1.3E-02  &  7.1E-03  &  4.9E-03  &  5.5E-02  &  5.3E-03  &  4.2E-03  &  9.2E-03  &  1.4E-02  &  4.1E-03  &  9.3E-03  &  8.6E-03  &  1.1E-02  &  9.6E-03  &  6.9E-03  &  4.3E-02  &  5.4E-03 \\
$m(N)/\text{\#R}$  &  0.19  &  0.03  &  0.20  &  0.57  &  0.72  &  0.22  &  0.19  &  0.07  &  0.73  &  0.07  &  0.11  &  0.19  &  0.37  &  0.09  &  0.37  &  0.24  &  0.29  &  0.29  &  0.19  &  1.22  &  0.15 \\
\hline
\end{tabular}
\end{adjustwidth}
\caption{For every class $C_m$, $m = 1, \ldots, 21$, the mean $\mu(N)$, median $m(N)$, maximum likelihood $\widehat{N}$, 90\% credible interval $(N_{90}^l, N_{90}^l)$, ratio $m(N)/\textsc{loc}$ median to lines of code, and ratio $m(N)/\text{\#R}$ median to number of methods of the distribution of total bugs in the class. The analysis started with $\widehat{\alpha}, \widehat{\beta}$ given by the maximum likelihood for $\alpha, \beta$ in $m[\alpha, \beta]$ computed from a \emph{uniform} prior.}
\label{tab:nbugs:uniform}
\end{table}

\begin{table}[!h]
\centering
\scriptsize
\setlength{\tabcolsep}{1pt}
\begin{adjustwidth}{-7mm}{-10mm}
\begin{tabular}{l| *{21}{r}}
& $C_1$  & $C_2$  & $C_3$  & $C_4$  & $C_5$  & $C_6$  & $C_7$  & $C_8$  & $C_9$  & $C_{10}$ & $C_{11}$ & $C_{12}$ & $C_{13}$ & $C_{14}$ & $C_{15}$ & $C_{16}$ & $C_{17}$ & $C_{18}$ & $C_{19}$ & $C_{20}$ & $C_{21}$ \\
\hline
$\mu(N)$  &  13  &  1  &  7  &  32  &  34  &  7  &  7  &  1  &  7  &  7  &  7  &  1  &  36  &  1  &  10  &  10  &  1  &  41  &  1  &  70  &  1 \\
$m(N)$  &  23  &  0  &  5  &  28  &  54  &  19  &  3  &  0  &  3  &  6  &  5  &  0  &  22  &  0  &  8  &  8  &  0  &  23  &  0  &  74  &  0 \\
$\widehat{N}$  &  9  &  0  &  4  &  26  &  26  &  4  &  4  &  0  &  4  &  4  &  4  &  0  &  26  &  0  &  6  &  6  &  0  &  40  &  0  &  60  &  0 \\
$N_{90}^l$  &  5  &  0  &  2  &  18  &  20  &  2  &  2  &  0  &  2  &  2  &  2  &  0  &  22  &  0  &  4  &  4  &  0  &  25  &  0  &  50  &  0 \\
$N_{90}^h$  &  26  &  6  &  17  &  56  &  59  &  17  &  17  &  6  &  17  &  17  &  17  &  6  &  62  &  6  &  21  &  21  &  6  &  68  &  6  &  105  &  6 \\
$m(N)/\textsc{loc}$  &  2.7E-02  &  0.0E+00  &  8.6E-03  &  1.4E-02  &  4.8E-02  &  3.4E-02  &  2.0E-03  &  0.0E+00  &  1.9E-02  &  1.2E-02  &  2.6E-03  &  0.0E+00  &  1.0E-02  &  0.0E+00  &  7.3E-03  &  3.9E-03  &  0.0E+00  &  8.4E-03  &  0.0E+00  &  2.9E-02  &  0.0E+00 \\
$m(N)/\text{\#R}$  &  0.43  &  0.00  &  0.14  &  0.57  &  0.84  &  0.59  &  0.05  &  0.00  &  0.25  &  0.15  &  0.07  &  0.00  &  0.26  &  0.00  &  0.29  &  0.11  &  0.00  &  0.25  &  0.00  &  0.83  &  0.00 \\
\hline
\end{tabular}
\end{adjustwidth}
\caption{For every class $C_m$, $m = 1, \ldots, 21$, the mean $\mu(N)$, median $m(N)$, maximum likelihood $\widehat{N}$, 90\% credible interval $(N_{90}^l, N_{90}^l)$, ratio $m(N)/\textsc{loc}$ median to lines of code, and ratio $m(N)/\text{\#R}$ median to number of methods of the distribution of total bugs in the class. The analysis started with $\widehat{\alpha}, \widehat{\beta}$ given by the maximum likelihood for $\alpha, \beta$ in $m[\alpha, \beta]$ computed from \emph{Jeffreys} prior.}
\label{tab:nbugs:jeffreys}
\end{table}

\PlotsForPriors{marginals}{Marginals $m[\alpha]$ and $m[\beta]$ of the posterior distribution $m[\alpha, \beta]$ of parameters $\alpha$ and $\beta$ for uniform priors (left) and for Jeffreys prior (right).}

\PlotsForPriors{posts}{Cumulative distribution function $W_{\alpha, \beta}$ for the values of $\alpha$ and $\beta$ in \autoref{fig:marginals-all-priors} for uniform priors (left) and for Jeffreys prior (right).}

\PlotsForPriors{derived_0}{Marginals of the probability distribution for a module having zero bugs detected by random testing with strong specifications for the values of $\alpha$ and $\beta$ in a 90\% credible interval for uniform priors (left) and for Jeffreys prior (right).}

\PlotsForPriors{derived_5}{Marginals of the probability distribution for a module having at most five bugs detected by random testing with strong specifications for the values of $\alpha$ and $\beta$ in a 90\% credible interval for uniform priors (left) and for Jeffreys prior (right).}

\GraphNBugsPrior{nbugs}{uniform}{Distributions $N_m$ of total bugs in module $C_m$ (starting with uniform priors to get values of $\widehat{\alpha}, \widehat{\beta}$ for prior Weibull).}


\fi

\end{document}